\documentclass{emulateapj} 
\usepackage{apjfonts}
\usepackage{epsfig}
\usepackage{xspace}
\usepackage{natbib}
\usepackage{hhline}
\usepackage{amsmath}
\usepackage{multirow}
\usepackage{dcolumn}
\usepackage{verbatim}
\usepackage{ulem}
\usepackage{rotating}

\newcolumntype{.}{D{.}{.}{-1}}

\newcommand{\chandra}{{\it Chandra}\xspace}
\newcommand{\champlane}{ChaMPlane\xspace}

\newcommand{\wavdetect}{{\it wavdetect}\xspace}

\newcommand{\spitzer}{{\it Spitzer}\xspace}

\newcommand{\sS}[1]{\mbox{$\rm{}^{#1}$}}
\newcommand{\Ss}[1]{\mbox{$\rm{}_{#1}$}}

\newcommand{\Ms}{\mbox{$M_{\odot}$}\xspace}
\newcommand{\Msi}{\mbox{$M_{\odot}^{-1}$}\xspace}

\newcommand{\nH}{\mbox{$N_{\mbox{\scriptsize H}}$}\xspace}
\newcommand{\nHt}{\mbox{$N_{\mbox{\scriptsize H22}}$}\xspace}

\newcommand{\Sx}{\mbox{$S_X$}\xspace}
\newcommand{\Hx}{\mbox{$H_X$}\xspace}
\newcommand{\Bx}{\mbox{$B_X$}\xspace}
\newcommand{\Sc}{\mbox{$S_C$}\xspace}
\newcommand{\Hc}{\mbox{$H_C$}\xspace}
\newcommand{\Bc}{\mbox{$B_C$}\xspace}
\newcommand{\Gs}{\mbox{$G1$}\xspace}
\newcommand{\Gm}{\mbox{$G2$}\xspace}
\newcommand{\Gh}{\mbox{$G3$}\xspace}

\newcommand{\lnls}{{log$N$-log$S$}\xspace}
\newcommand{\Deg}{\mbox{$^\circ$}\xspace}
\newcommand{\Arcmin}{\mbox{$\arcmin$}\xspace}

\newcommand{\ce}{\mbox{$\sigma_{s}$}\xspace}
\newcommand{\se}{\mbox{$\sigma_{c}$}\xspace}

\newcommand{\nHu}{\mbox{$\times 10^{22}$ cm\sS{-2}}\xspace}
\newcommand{\lcgs}{\mbox{ergs s\sS{-1}}\xspace}
\newcommand{\fcgs}{\mbox{ergs cm\sS{-2} s\sS{-1}}\xspace}
\newcommand{\pcgs}{\mbox{ph cm\sS{-2} s\sS{-1}}\xspace}

\begin{document}

\title{Radial Distribution of X-ray Point Sources
near the Galactic Center}


\author{
JaeSub Hong\altaffilmark{1*},
Maureen van den Berg\altaffilmark{1},
Jonathan E. Grindlay\altaffilmark{1},  
and 
Silas Laycock\altaffilmark{2}
}
\altaffiltext{*}{Send requests to J. Hong at jaesub@head.cfa.harvard.edu}
\altaffiltext{1}{Harvard-Smithsonian Center for Astrophysics, 
60 Garden St., Cambridge, MA 02138 }
\altaffiltext{2}{Gemini Observatory, 670 N. A'ohoku Place,
Hilo, HI 96720}

\begin{abstract} 

We present the \lnls and spatial distributions of  X-ray point sources in
seven Galactic Bulge (GB) fields within 4\Deg from the Galactic Center
(GC). We compare the properties of 1159 X-ray point sources discovered
in our deep (100 ks) \chandra observations of three low extinction Window
fields near the GC with the X-ray sources in the other GB fields centered
around Sgr B2, Sgr C, the Arches Cluster and Sgr A* using \chandra
archival data.  To reduce the systematic errors induced by the uncertain
X-ray spectra of the sources coupled with field-and-distance dependent
extinction, we classify the X-ray sources using quantile analysis and
estimate their fluxes accordingly.  The result indicates the GB X-ray
population is highly concentrated at the center, more heavily than the
stellar distribution models.  It extends out to more than 1.4\Deg from
the GC, and the projected density follows an empirical radial relation
inversely proportional to the offset from the GC.  We also compare
the total X-ray and infrared surface brightness using the \chandra and
\spitzer observations of the regions.  The radial distribution of the
total infrared surface brightness from the 3.6 band $\mu$m images appears
to resemble the radial distribution of the X-ray point sources better than
predicted by the stellar distribution models.  Assuming a simple power
law model for the X-ray spectra, the closer to the GC the intrinsically
harder the X-ray spectra appear, but adding an iron emission line at
6.7 keV in the model allows the spectra of the GB X-ray sources to be
largely consistent across the region.  This implies that the majority of
these GB X-ray sources can be of the same or similar type.   Their X-ray
luminosity and spectral properties support the idea that the most likely
candidate is magnetic cataclysmic variables (CVs), primarily intermediate
polars (IPs). Their observed number density is also consistent with the
majority being IPs, provided the relative CV to star density in the GB
is not smaller than the value in the local solar neighborhood.

\end{abstract}


\section{Introduction}

The \chandra X-ray Observatory has opened a new era in studies of the
X-ray source population in the Galactic Bulge (GB). A series of shallow
and deep \chandra observations in the Galactic Center (GC) region have
revealed $\sim$ 1000 X-ray point sources in a 2\Deg $\times$ 0.8\Deg region
\citep{Wang02} and 2357 X-ray point sources in a $17\Arcmin \times
17\Arcmin$ region around the Sgr A* \citep[hereafter M03]{Muno03}.
An additional $\sim$ 2000 sources found in the Bulge Latitude Survey
(BLS, two 0.8\Deg $\times$ 1.5\Deg regions) provide the initial results
for the latitude
distribution of the GB sources \citep{Grindlay09}.  The X-ray luminosities
and relatively hard spectra ruled out that the majority of the GC X-ray
point sources are normal stars, active binaries, young stellar objects,
or quiescent low mass X-ray binaries (qLMXBs) (M03).  From the lack of
real matches between the bright infrared (IR, $K<15$) and X-ray
sources in the Sgr A* field, \citet[hereafter L05]{Laycock05} 
concluded that high mass X-ray binaries (HMXBs) cannot account for
more than  10\% of the X-ray sources in this region.  While the
leading candidate that fits the properties of these X-ray
sources is now magnetic cataclysmic variables (CVs) \citep[L05]{Muno04},
the relatively hard X-ray spectra of some of the most recently
discovered qLMXBs imply qLMXBs could be misrecognized as CVs and be
more common in the GB than thought in the past \citep{Wijnands05,
Bogdanov05}.  Infrared (IR) searches for the counterparts of these GB
X-ray sources have been actively pursued (e.g.  \citet{Muno05}), but
the exact nature of the majority of the sources is still elusive due
to high obscuration by dust and source confusion by the high star density.  

We have conducted a series of deep (100 ks) \chandra observations
of three low extinction Window fields -- Baade's Window (BW), Stanek's
Window (SW; Stanek 1998) and the "Limiting Window" (LW) -- near the GC (\S\ref{s:field}).
These Window fields allow us to observe the GB X-ray population and
their Galactic radial distribution with minimal obscuration by dust.
We have discovered 1159 X-ray point sources in these fields. 
We compare
their distributions with X-ray sources in other GB fields -- the Sgr B2,
Sgr C, Arches Cluster and Sgr A* fields.  We present a new approach using
quantile analysis (\S\ref{s:qa}) to minimize the systematic errors in flux estimation, 
to classify sources by their X-ray spectral types and investigate
their radial distribution.  We compare the X-ray distribution
with the known models of the stellar distribution (\S\ref{s:dist}) and investigate
the nature of the X-ray population (\S\ref{s:disc}). 
See also \citet{Berg09}, where we put constraints on the nature of
the X-ray source populations from the optical point of view, using HST
observations of the Window fields taken simultaneously with the \chandra
exposures.  This work is part of our \chandra Multi-wavelength Plane
(\champlane) Survey designed to measure the space density and probable
nature of the low-luminosity accretion-powered sources in the Galaxy
\citep{Grindlay05}.

\setcounter{footnote}{1}
\section{Observations and Data Analysis} \label{s:field}

We performed \chandra/ACIS-I observations of BW on 2003 July 9
(Obs.~ID 3780), SW on 2004 February 14/15 (Obs.~ID 4547 and 5303), and
LW on 2005 August 19/22 and October 25 (Obs.~ID 5934, 6362 and 6365). 
Due to technical constraints, the SW and LW observations were
segmented into a few pointings, which we stacked for further
analysis.  Table \ref{t:f} summarizes the observational parameters and
X-ray source statistics of the Window and other GB fields analyzed in
this paper.  For the Sgr A* field, we use the results from a 100 ks
observation (Obs.~ID 3665) for easy comparison with other GB fields
that were observed with similar exposure times, and we have also
stacked 14 observations from the archive, totaling 750 ks exposure.

We have analyzed the data as a part of the \champlane survey. For uniform
analysis of all the \champlane fields, we have developed a series of
X-ray processing tools, mainly based on version 3.4 of the CIAO package
\citep[hereafter H05]{Hong05}\footnote{Some of the fields were
processed by the tools based on version 3.1 of the CIAO package,
but the difference between two versions is minimal.}.  After initial
screening of the CXC level-2 data (e.g.  select the events in good
time intervals during which the background fluctuates $<3\sigma$ above
the mean level), we detect X-ray point sources with a wavelet
algorithm (\wavdetect, \citealt{Freeman02}) with a significance
threshold of $10^{-6}$. The \wavdetect routine is run on each
individual observation and the stacked data set if available. Multiple
observations are considered stackable (the SW, LW and Sgr A* fields
here) if the aimpoints are on the same detector (ACIS-I or ACIS-S) and
they are within 1\Arcmin of each other. 

In H05, we used source detections in the broad (\Bx: 0.3--8.0 keV)
band. We now also incorporate source detections in the soft (\Sx: 0.3--2.5
keV) and hard (\Hx: 2.5--8.0 keV) bands in addition to the broad band.
We establish a unique source list  by cross matching the three detection
lists based on the relative distance of possibly identical source pairs
(the closest pairs) in the three different bands. The relative distance
($d_r$) of two sources is defined by the ratio of the source distance
to the quadratic sum of the positional errors.  Note that there is no
astrometric offset among the images in the three bands. The positional errors
of sources are calculated by an empirical formula based on the 
MARX\footnote{http://space.mit.edu/CXC/MARX/.} simulations (Eq.~5 in H05).

\begin{figure*}[H] \begin{center} 
\plotone{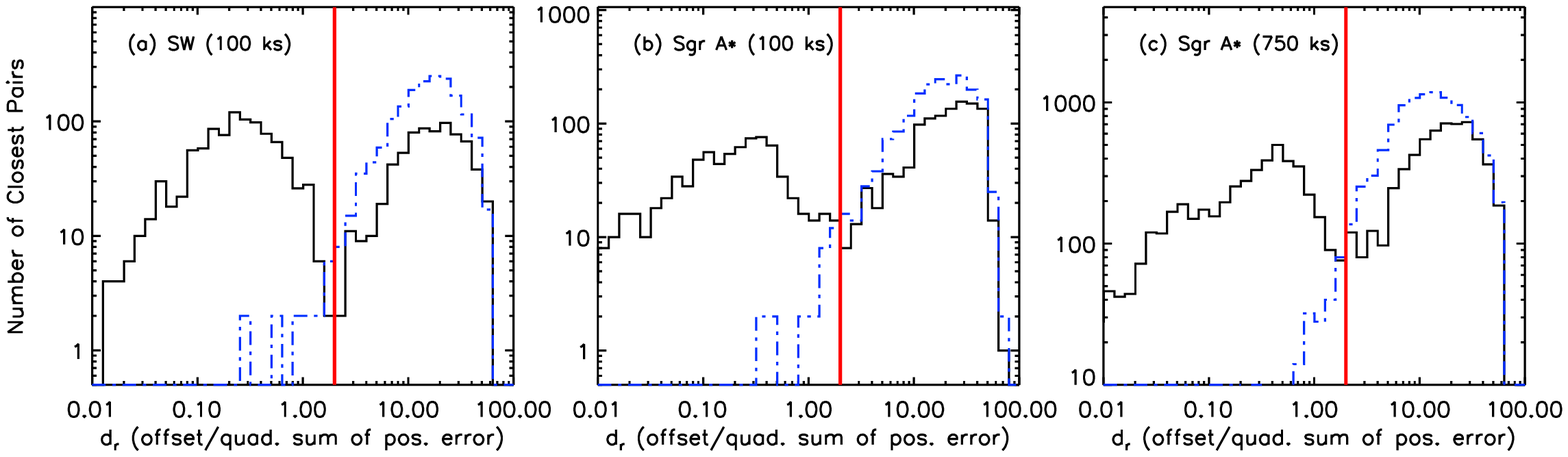} 
\end{center} 
\caption{Cross correlating X-ray sources detected in three different
energy bands: the number of pairs as a function of the
relative separation ($d_r$, see text) of potentially identical sources
(closest pairs) among the three detection bands.  The plots show the results from
the 100 ks observation of the Stanek Window (a), the Sgr A* field (b)
and 750 ks of the Sgr A* field (c).  The bimodal shape is due to the
mixture of the true and random matches in the distribution.
The (blue) dashed-dotted lines show the results after introducing
an arbitary global offset (1\Arcmin in both R.A. and Dec.) among the
three band detections, which illustrate the distributions of the random
matches.  
}
\label{f:xm} \end{figure*}

Establishing a unique source list is straightforward in relatively
un-crowded fields such as the Window fields, but it can
be tricky in heavily populated fields and in very deep exposures 
such as the stacked Sgr A* dataset.
Fig.~\ref{f:xm} shows the distributions of the relative
separations of nearest-neighbor source pairs among the three detection
bands, \Sx, \Hx, and \Bx.
As examples, we compare the 100 ks observations of the Stanek Window
and Sgr A* field, and the 750 ks stacked dataset of the Sgr A* field. 
A source detection
in each band contributes two pairs to the distribution, one from each
of the other two bands.

The bimodal shape of the distribution indicates two types of the
pairs contribute to the distribution: one type consists of the truly
identical sources detected in different bands and the other consists
of random pairs of unrelated sources. The distribution of the random
pairs can be estimated by introducing an arbitrary astrometric offset
in the source position between the detection bands. The (blue)
dashed-dotted line in Fig \ref{f:xm}  shows such an example (1\Arcmin offset in both R.A. and
Dec.), the shape of which closely resembles 
the right side of the bimodal distribution of original sources.
The slight excess over the original
distribution is due to the real pairs being transformed into new random
pairs by the positional offset.

After visual inspection of the raw images and the distributions of the
relative distances in Fig.~\ref{f:xm}, we use a simple cut ($d_r\le$2.0,
red vertical line) for establishing the unique source list.  The cut is
sufficient for identifying virtually 100\% of the unique sources detected
in multiple bands. From the distributions of the relative distances of
the random pairs, we estimate that the false random matches surviving the
cut ranges from 10 to 25 for the 100 ks observations.  The corresponding
number of independent sources that might have lost by the false random matches ranges
from 5 to 10 ($\sim$ 1\%) for the 100 ks observations and about 100 for
the stacked Sgr A* field ($\sim$ 3\%).

When multiple detections in the three bands are identified as a unique
source, we select the one with the smallest positional error for the
unique source list.  We note that the final source position and error
could be derived from a form of weighted average of astrometric properties
of multiple detections.  However, the \Sx and \Hx band detections are not entirely
independent from the \Bx band detection. Therefore, in order to avoid
unnecessary complication in the analysis, we simply take the astrometric
(and photometric) properties of the source with the smallest positional
errors, which is the detection with the highest significance
among the three bands.

As a sanity check, we compare our detection with the available catalogues
in the literature.  M03 provided a catalogue of the 2357 X-ray point
sources discovered in the 690 ks exposure (626 ks of GTIs) of the Sgr A*
field,   \citet[hereafter M06]{Muno06} for the 397 point sources in
the 100 ks exposure of the Sgr B2 field (Obs. ID 944) and
\citet[hereafter W06]{Wang06} for the 244 X-ray point sources in the 100
ks exposure of the Arches Cluster (Obs. ID 4500).  The majority ($\sim
85 - 90$\%) of these sources are also detected in our analysis or vice
versa if our source list is shorter than theirs.  A small fraction of
the sources are missing due to many subtle differences in the detection
methods and the selection criteria for the lists such as the detection
energy bands (0.5--8, 0.5--1.5 and 4--8 keV in M03 or 1--4, 4--9 and
1--9 keV in W06), the pointing or GTI selections (e.g. Obs.~ID 1561 was
not included in our analysis in the stacked Sgr A* field). In addition,
in the case of W06, their high
significance threshold (10\sS{-7} except for the inner $2\Arcmin \times 2\Arcmin$ region, 
about 10 counts in their broad band, 1--9 keV) 
is one of the main reasons for the difference (244 vs.~423) in the 
total number of the source detections. M03 and M06 list all the source detections 
including ones with negative counts in their broad band (0.5--8 keV).
Our source list includes the detections with net counts $\ge$ 1 in the \Bx band 
(0.3--8 keV).  

After source detection, we perform aperture photometry on each source to
extract the basic source properties such as net count and net count rate
in the conventional energy bands (\Sc: 0.5--2.0, \Hc: 2.0--8.0 and \Bc:
0.5--8.0 keV) and energy quantiles in the broad band (\Bx: 0.3--8.0 keV).
For the sources that fall near other sources, we carefully revise the
aperture of the source regions by excluding overlapping sections to
minimize the contamination
from the neighbors (H05). Table \ref{t:sc} lists a part of the source
catalog with selected source properties used in this paper. The complete
list for the Window and other GB fields is available in the electronic
edition.

\section{Flux estimation by Quantile classification} \label{s:qa}


In order to compare source distribution in various regions of the
sky with diverse extinctions, it is necessary to correct for the
interstellar absorption and use the unabsorbed source flux of individual
sources. However, such a calculation is not trivial for X-ray sources with
diverse spectral types as found in Galactic plane fields since faint sources
are unsuitable for spectral fitting.  Moreover, the relatively large
extinction in the GB fields and its usually-unknown field-and-distance
dependent variation make it difficult to identify the underlying
X-ray spectral model (e.g. power law vs. thermal Bremsstrahlung, etc).
An inaccurate assumption of the spectral model when estimating flux
introduces systematic errors that often exceed the statistical errors.
We therefore employ quantile analysis \citep[hereafter H04]{Hong04},
which is relatively free of the count-dependent bias inherent in X-ray
hardness ratio or X-ray color analysis, and so provides a better measure
for classifying X-ray sources in the GB fields. In the following,
the energy quantile $E_x$ corresponds to the energy below which $x\%$
of the counts are detected.  e.g. $E_{50}$ is the median energy.

\subsection{Quantile Analysis}

\begin{figure*}[t] 
\plotone{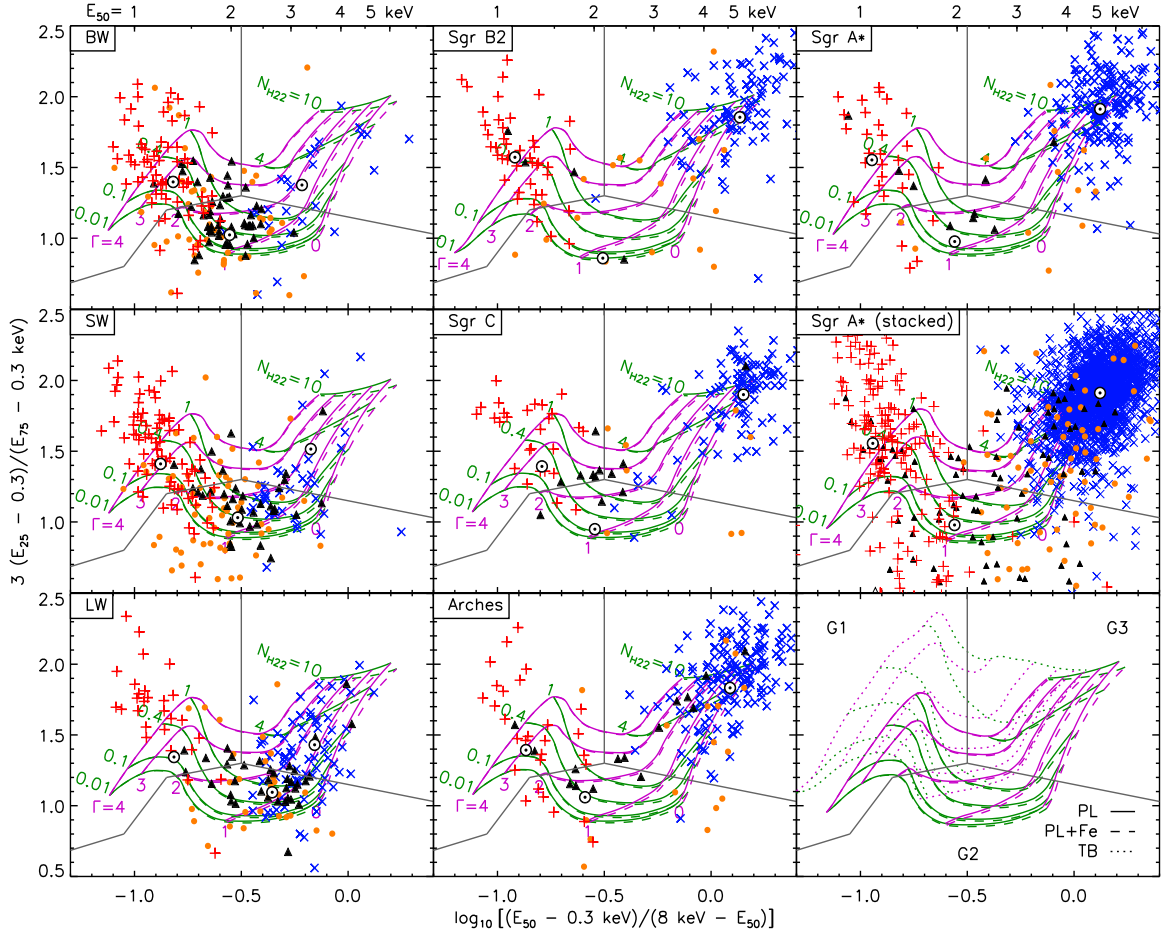}
\caption{Quantile diagrams (0.3--8 keV) of the X-ray sources with $S/N
\ge 3$ in the GB fields overlaid with grids for a simple power law
model (PL, solid lines, power law index $\Gamma$ = 0, 1, 2, 3 \& 4,
\nHt = 0.01, 0.1, 0.4, 1 \& 10), a power law plus an iron line model
(PL+Fe, dashed, at 6.7 keV with 0.4 keV EW), and thermal
Bremsstrahlung model (TB, dotted, $kT$ = 0.2, 0.4, 1, 2, 4 \& 10 keV,
\nHt = 0.01, 0.1, 0.4, 1 \& 10, only shown in the bottom-right plot
for clarity). The energy quantile $E_x$ corresponds to the energy
below which $x\%$ of the counts are detected.  The (red) crosses are
for the relatively soft sources ($S/N \ge 3$ in \Sc, but not in \Hc),
the (blue) `x's for the hard sources ($S/N \ge 3$ in \Hc, but not in
\Sc), the (black) triangles for the bright sources ($S/N \ge 3$ in
both \Sc and \Hc), and the (orange) dots for the faint sources ($S/N
\ge 3$ only in \Bc).  The (grey) lines from ($-$0.5,1.3) divide
each diagram into the soft (\Gs), medium (\Gm) and hard groups (\Gh).  
$\odot$s mark the quantiles of stacked
photons in each group. 
}
\label{f:qd} 
\end{figure*}

Fig.~\ref{f:qd} shows the quantile diagrams of the X-ray sources ($S/N
\ge 3$ in \Bx) in the selected GB fields overlaid with grids for a
simple power law model (PL, solid lines) and a power law plus an iron
 emission line (PL+Fe, dashed) at 6.7 keV with 0.4 keV equivalent width
(EW, see \S\ref{s:hard} for the motivation of the line choice). The lower right
panel also includes a grid for the thermal Bremsstrahlung model (TB,
dotted).  The $S/N$ here is calculated based on the statistical errors
(\se) using small-number statistics from \citet{Gehrels86} (see also \citet{Kim04}).
The difference in the model grids between PL and PL+Fe is only evident
in the highly absorbed or spectrally very hard section of the diagram (the right
side) because a small iron line ($>$ 6 keV with $<$ 1 keV EW) does not
make a noticeable difference in three quantiles of the soft sources.

Relatively insensitive to the extinction, the sources around $(x,y) =
(-0.9,1.6)$ are present in every field and they appear unabsorbed
and intrinsically soft regardless of the assumed model class (PL,
PL+Fe or TB).  Foreground thermal sources such as coronally active
stars fit the description.  The location of relatively hard sources in
the diagram varies with the field extinction.  In the Window fields,
the hard sources are relatively unabsorbed, but on approach to the GC,
there is an increasing trend in the source number with both the average
absorption and the intrinsic hardness, when compared to a simple power
law model. For instance, in BW most sources have power law photon index
($\Gamma) >1$ and $\nHt<1$, whereas in LW many sources lie in $\Gamma
<1$ and $\nHt>1$.  In the Sgr A* field and the rest, most of the hard
sources are heavily absorbed with $\Gamma \lesssim 1$  and $\nHt\gtrsim
1$ on average, well separately from the foreground sources. We explore
this more in Section 3.2.

The quantile diagrams nicely illustrate the spectral diversity of
the X-ray sources in the GB fields, but poor photon statistics also
contributes to the scatter.  To reduce systematic errors caused
by poor statistics in assigning spectral types while allowing the
spectral diversity of the sources in each field, we divide the diagram
into three groups as shown by the (grey) lines originating at ($-$0.5,
1.3).  The left section represents most foreground thermal sources (\Gs:
soft group), the middle section most unabsorbed accreting sources (\Gm:
medium group), and the right section the absorbed thermal or accreting
sources (\Gh: hard and absorbed group).  The division between \Gs and
\Gh is devised to be somewhat robust\footnote{Under the PL model, the
boundary of the \Gs and \Gm groups stays in roughly between $\Gamma$=2 and 3.}
against variations in detector response between ACIS-I and ACIS-S (see
H04; H05); or induced by gradual loss of low energy response.  The final results
(e.g. \lnls distributions) are not sensitive to small changes of the
group boundaries (e.g. shifting the boundaries by $\sim$ 0.1 in $x$ or
$y$).  The mean quantiles for each group (marked by $\odot$) are
calculated by the stacked photons of the sources in the group with
$S/N \ge 3$ and net counts $\le 1000$ (to avoid being dominated by a
few bright sources) in \Bx.  For a given model class (e.g. PL), we
estimate the spectral model parameters (e.g. $\Gamma$ and \nH) of the
sources in each group using the mean quantiles.

\subsection{Spectral hardening vs. radial offset from GC} \label{s:hard}

Table \ref{t:q} summarizes the group mean quantiles of the \Gm and \Gh
sources and corresponding model parameters under the PL and PL+Fe models.
The \Gm sources in the
high extinction fields are omitted in the Table since they are mostly foreground sources. 
For comparison, the table also shows the model parameters estimated from
spectral model fits. In order to increase photon statistics, we stacked the
spectra of sources within a group with net counts $\le 1000$ for spectral
fitting\footnote{One can use a spectral model fit on individual sources
with net counts $>$ 200--300, but in order to establish more reliable
statistics for the presence of the line emission for the group, we also stack
moderately bright sources with net counts up to 1000.} and the spectra were binned to have
at least 40 counts in each bin.  The model used is a power law plus an
iron emission line for which we have chosen the 6.7 keV Fe XXV He-$\alpha$
line because it has also been observed in the spectra of the X-ray
point sources in  the deep survey of the Sgr A* field (M03),
the shallow survey of the GC strip \citep{Wang02}, and other parts of
the Galactic plane \citep{Ebisawa05}.  The 6.4 keV neutral iron line
is also present in some sources of the GB fields, but it is generally
more prominent as unresolved diffuse emission \citep{Wang02}. Note our
aperture photometry is designed to minimize possible contamination of
the diffuse emission through background subtraction (H05).  We have
chosen the 0.4 keV EW for the line in quantile analysis (and the \lnls
distribution later) because it lies in the EW range
estimated by spectral model fits on the \Gh sources and
\citet{Muno04} found a similar value ($\sim$ 0.4 keV) for the bright
sources in the Sgr A* field.

Under the PL model, both groups show a trend of increasing hardness of
the intrinsic spectra on approach to GC (i.e. $\Gamma \gtrsim 1$ for
the Window fields vs.~$\Gamma < 1$ for other GB fields). 
This apparent trend can be attributed to a few factors.  

In the case of the \Gm sources
in the Window fields, the group contains a large number of foreground
coronal sources and background AGN in addition to the GB sources.
For instance, in the BW and SW, about 70\% of the sources with $S/N \ge 3$
in the \Hc band are background AGN (see Fig.~\ref{f:lnls} in \S\ref{s:dist}).
Therefore, the group is perhaps too contaminated for the apparent
trend to be taken for real.

\begin{figure}[t] 
\plotone{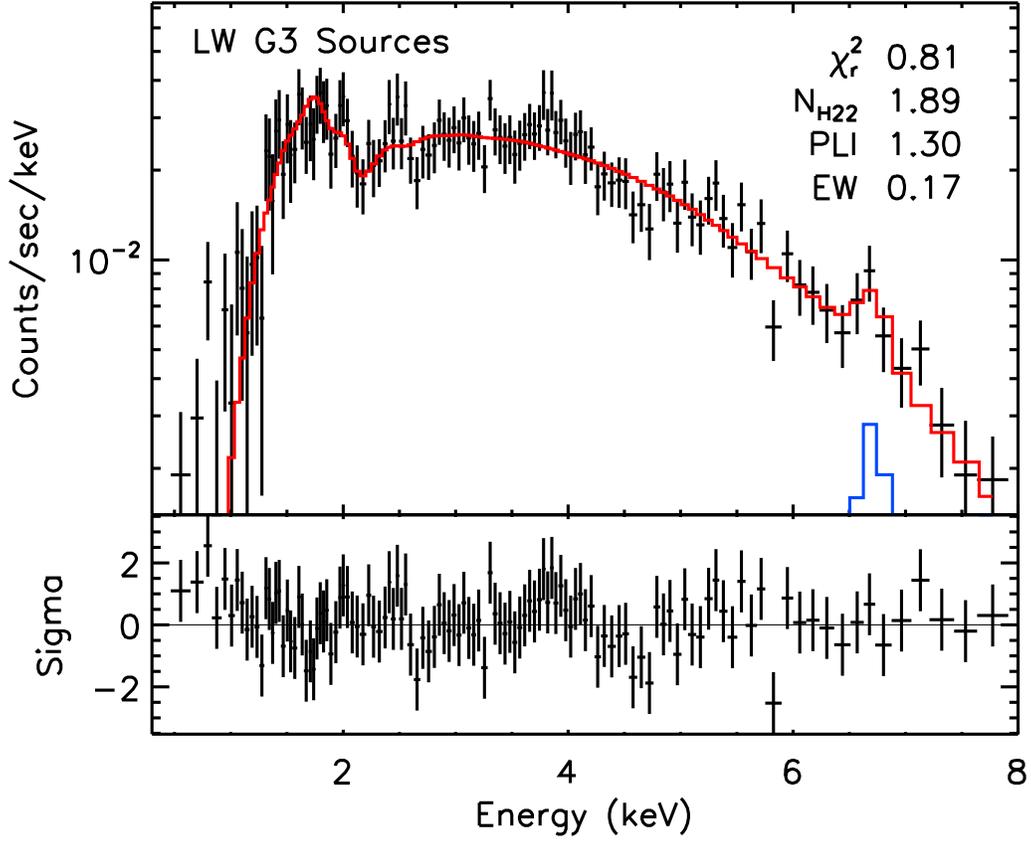}
\caption{
The stacked spectra of the \Gh sources (net counts $<$ 1000)
in LW with the PL+Fe XXV He-$\alpha$ (6.7 keV line) fit. The estimated EW
of the line is $0.17 \pm 0.08$ keV. 
}
\label{f:lw} 
\end{figure}

The similar trend in the \Gh group  appears to be more realistic. However,
comparing the PL+Fe model grid with the PL model grid in the quantile
diagram suggests this trend can be an artifact of using the PL model at
least in part.  Indeed the trend is alleviated under the PL+Fe model
as shown by the model parameters estimated by both quantile analysis and
the spectral fits (Table \ref{t:q}).  Poor statistics in the \Gh
sources of BW and SW does not allow any meaningful constraint of the
iron line emission in the spectra, but the stacked spectrum of the \Gh
sources in the LW does show a clear hint of the 6.7 keV line
(Fig.~\ref{f:lw}, see also \citet{Revnivtsev09}).  The relatively weak
line feature (EW $\sim$ 0.17 keV) in the LW can be explained by the
relatively large contribution
of AGN in the group compared to other GB fields (see Fig.~\ref{f:lnls}
in \S\ref{s:dist}).  

With the inclusion of a 6.7 keV line,  the power law index ($\Gamma$)
becomes largely consistent across the GB fields with the possible exception
of the Sgr B2 or C field (see \S\ref{s:sgrc}).  The result indicates that the galactic
X-ray sources in the LW field may be the same type of sources as seen in
the other GB fields closer to the GC. If true, the GB X-ray sources indeed
extend out to at least 1.4\Deg from the GC.

\subsection{Flux Estimates}

Based on the group model parameters, we estimate the source flux in
the conventional energy bands, using the instrument response files at
the aim point and scaling them for source position on the detector by
the exposure map (H05)\footnote{The latest CIAO tools (ver 3.4 or
higher) can calculate the response files appropriate for each source
location.}.  For the stacked data, we use the average response files
weighted by the exposure of each observation.

\begin{figure*}[t] \begin{center} 
\plotone{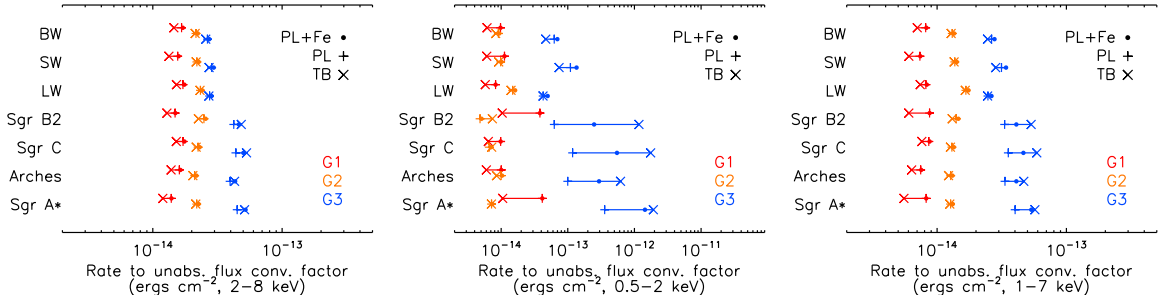} 
\end{center} 
\caption{Comparison of the rate-to-flux conversion factor 
for the three quantile groups of sources in the three energy ranges.
The conversion factor in the \Hc band (2--8 keV) is robust ($\lesssim$
20--30\% variation), in the \Sc band (0.5--2 keV) it is very unreliable (up to
more than a factor of 10), and in the medium energy range (1--7 keV),
there are significant variations (up to $\sim$ 100\%).
}

\label{f:rfc} \end{figure*}

The quantile diagram can assign the model parameters (e.g.
$\Gamma$=1.7 vs.~1.0) appropriate to a given model class for the
sources, but it cannot determine which model class (e.g. PL vs.~TB) is
right for the sources. A certain model can only be ruled out when the
derived values of the parameters are unphysical or with external information (e.g. optical identifications).  In order to estimate
the systematic errors arising from the improper choice of the model
class, we compare flux estimates under three different model classes:
PL, PL+Fe and TB.  In order to see the significance of the difference
among these models, Fig.~\ref{f:rfc} compares the conversion factor of
count rate to unabsorbed flux for sources near the aim point in each
group under the three model classes.  In the case of the \Hc band, the
difference between the model classes is very small, but in the \Sc
band, the conversion factor can differ by more than a factor of 10. 
We take the largest difference in the flux estimates among the three
model classes as the systematic errors (\ce) and compare them with the
pure count-based statistical errors (\se).  For the flux estimates in
the \Hc band, $\ce \sim 20 - 30\%$ and we get  $\ce < \se$ for $\sim$
87\% of the sources with $S/N \ge 3$, and $\ce < \se$ for $\sim$ 62\%
even with $S/N \ge 5$.  In the \Sc band, $\ce$ can be larger than
1000\%, and we get $\ce < \se$ only for 28\% with $S/N \ge 3$ and for
16\% with $S/N \ge 5$.  

This exercise does not explore all the possible model classes, but the
results indicate the \Hc band flux estimates in this method are robust
and relatively insensitive to the choice of the model classes. 
However, the \Sc band flux estimates can be dominated by the
systematic errors arising from improper selection of the spectral
model.  The fundamental difference between the \Sc and \Hc band is
that the \Sc band is very sensitive to the range of interstellar
absorption in the GB fields, $\sim 10^{21-23}$ cm\sS{-2}, while the
\Hc band is not. In the following, we limit our discussion to the \Hc
band results using the sources with $S/N \ge 3$ in \Hc (`x's and
triangles in Fig.~\ref{f:qd}), which are likely to be GB sources (and
AGN) rather than the foreground sources.  See also \citet{Berg09} for
the spectral choices for the flux estimates in \Bx.  If the number of
counts in the stacked spectrum of a quantile group (\Gs, \Gm or \Gh)
were large enough, a spectral fit would be better suited for
determining the underlying spectral model and its parameters, but a
fit can also leave ambiguity over the correct spectral model. 
Since in the \Hc band the difference driven by the model
class is less significant than the statistical errors, we simply use
the model parameters estimated by quantile analysis.


\section{source distribution} \label{s:dist}

\subsection{Eddington and Malmquist Biases} \label{s:bias}

In order to explore the effect of the Eddington bias (EB), which makes
the faintest sources appear brighter than they really are, we simulate three
spectral types of sources (one for each group) based on the group mean
quantiles for each field.  Using the MARX simulation code,
we generate the sources with net counts (\Bx) from 5 to 400,  using a
$S^{-1}$ distribution to cover the wide count range
efficiently.  We scatter 200 -- 250 of these sources randomly over the
real events and apply the regular analysis procedure.  We repeat the
procedure 1000 times. The fake sources are not allowed to overlap each
other but they can fall on top of the real sources.  The results indicate
that the EB is noticeable in the sources with $<$ 10 counts, which can appear
as bright as 15 -- 20 count sources, depending on the field (or up to $\sim$
30 count sources for the stacked Sgr A* field).  Since we consider sources
with $S/N \ge 3$ in the \Hc band, which corresponds to $\gtrsim$ 16 --
20 counts at least ($\gtrsim$ 30 counts for the stacked Sgr A* field),
we expect EB is not a major contributor to the errors of the following
distributions.

The Malmquist bias (MB) is due to the exposure dependent volume (depth)
coverage.  The MB is usually a concern for luminosity distributions
but not for \lnls distributions in the apparent (detected)
flux space.  However, the \lnls distributions in the unabsorbed flux space
can be subject to the MB when strong interstellar absorption 
limits the depth of the view, underestimating the true distribution.
Therefore, the faint end of the \lnls distribution can be lower than the true
distribution and so the MB counteracts the EB to some extent.

With $\ge$ 100 ks exposure, all the sources with an unabsorbed flux
$\gtrsim 10^{-14}$ \fcgs can be detected at the far
side of the Galaxy with $S/N \ge 3$ in \Hc under the assumption of
the total integrated absorption\footnote{
Assuming the absorption to the GC to be 6 \nHu \citep{Baganoff03} and
the symmetry with respect to the GC.} of \nH $\sim 12 \nHu$.  Therefore the MB is not a
concern for sources with $\gtrsim 10^{-14}$ \fcgs  (or $\gtrsim 5\times
10^{-15}$ \fcgs for the Window fields) under the assumption of a power
law wth $\Gamma$=1.0 for the X-ray spectrum.  This does not mean we
can access X-ray sources of a certain luminosity uniformly all the way
through the Galaxy. For instance, the unabsorbed flux of $S> 10^{-14}$
\fcgs corresponds to $L_X \gtrsim 8\times 10^{31}$ \lcgs  at the GC (8
kpc, \nH $\sim 6 \nHu$) and $L_X \gtrsim 7 \times 10^{32}$ \lcgs at 20
kpc (\nH $\sim 12 \nHu$).  The situation is a bit more tricky since
under the quantile group method we assign fixed spectral parameters
with a fixed \nH value for the X-ray spectra of all the sources in
each group, which in fact have a diverse \nH distribution (e.g. the
\Gh group in the Sgr A* field).  However, the sources with an
unabsorbed flux of $S> 10^{-14}$ \fcgs in this method are free of the MB
for $\gtrsim$ 100 ks exposure.  The MB can be a problem for the \Gs and \Gm
sources in the high extinction fields, but their contribution in the \lnls
distribution of the \Hc band is negligible compared to the \Gh sources.

\subsection{Sky Coverage} \label{s:skyc}

For the \lnls distribution, we need to know the sky coverage as a function
of flux.  In order to minimize the systematic errors associated with spectral
type assignment, we calculate the sky coverage of each source based on the
detected photon counts in the three energy bands as follows
(Cappelluti et al.~2005; H05).  For each observation, we generate the
background-only images by removing the counts in the source regions in
the image and filling the region with the counts using the statistics
in the surrounding regions ({\it
dmfilth})\footnote{http://cxc.harvard.edu/ciao/ahelp/dmfilth.html}. 
At every pixel in the background images, we calculate the minimum
source counts required for detection with $S/N\ge 3$.  For the sky
coverage of a given source, we take the sky area where the minimum
counts are less than the net counts of the source in the band.  These
sky coverage values agree well with those expected from the
simulated sources of three spectral types using the MARX
(\S\ref{s:bias}).  On average, they are within 10\% for the cases with
$S/N \ge 3$, which indicates this method accounts for the completeness as well.

\subsection{The \lnls and Radial Distributions} \label{s:lnls}

\begin{figure*}[t] 
\plotone{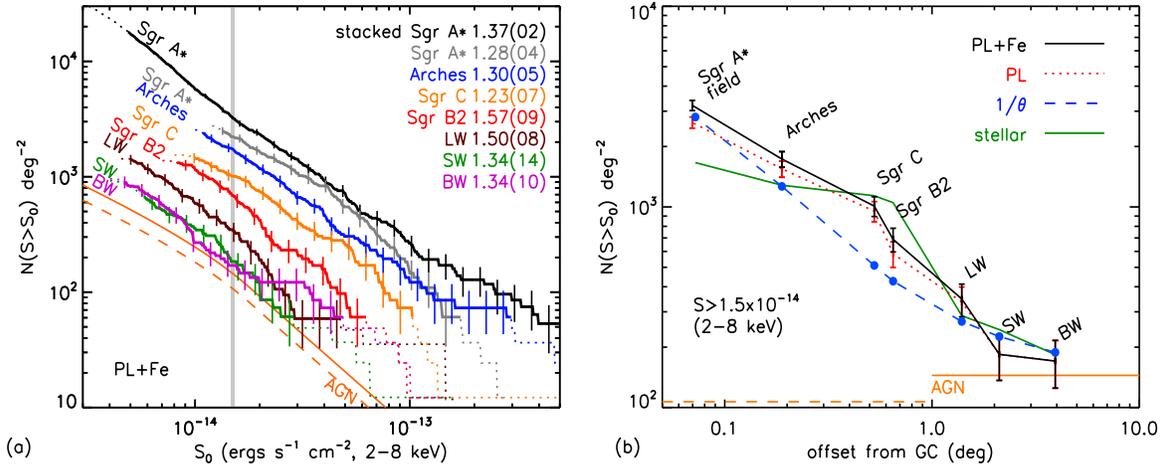}
\caption{The \lnls (a) and the radial (b) distributions of the
X-ray sources in the GB fields.  Fluxes are computed under the PL+Fe
model (6.7 keV line with 0.4 keV EW) and the distributions include the
sources with $S/N \ge 3$ in the \Hc (2--8 keV) band.  The numbers in
the legend of (a) are the slopes ($\alpha$) and their error
of the power law fits ($N\propto S^{-\alpha}$) to the solid section of
the \lnls distributions
(see the text for the definition of the solid section).
The (orange) solid line is the active galactic nuclei (AGN)
distribution from \citet{Kim07} seen in the low extinction fields and
the (orange) dashed line is the same corrected for the extinction to
the GC (\nH=6$\times10^{22}$ cm\sS{-2}, see the text). 
The radial distribution shows the number density of the X-ray sources
with $S> 1.5\times 10^{-14}$ \fcgs (marked by the vertical strip in
the left panel) under two spectral models (solid black for PL+Fe and
dotted red for PL), compared with the stellar distribution (solid green)
and the $1/\theta$ distribution (blue dashed). The $x$-axis in the
radial distribution is the average offset value of the sources in each field.  
}
\label{f:lnls} \end{figure*}

The \lnls and radial distributions of the X-ray
sources in the GB fields are shown in Fig.~\ref{f:lnls}a \& b respectively. 
The \lnls distribution was computed using sources with   $S/N\ge 3$ in
the \Hc band, under both the PL and PL+Fe models described in \S\ref{s:qa},
with the latter result plotted in Fig.~\ref{f:lnls}a.
The source number-density values plotted
against angular distance from Sgr A* in Fig.~\ref{f:lnls}b are
projected from the \lnls distributions at the flux value indicated by
the vertical grey line ($S> S_0 = 1.5 \times 10^{-14}$ \fcgs)
in Fig.~\ref{f:lnls}a.
As seen in Fig.~\ref{f:lnls}b,
the total source densities under the PL model are
slightly lower than the same under the PL+Fe model in the high extinction
fields, while both distributions under these models are nearly
identical in the three low extinction Window fields.

For clarity, we define the statistically robust section of each
distribution in Fig.~\ref{f:lnls}a and emphasize it with a solid line. 
This "solid section" is defined to contain
contributions from at least 10 or more sources, which
set the upper limit of the range (e.g. $S_0 \sim 10^{-13}$ \fcgs for
Sgr C).  The lower limit is set by the flux value at which the sky
coverage of the contributing \Gh sources is greater than 50\% of the
maximum sky coverage i.e. the full field of view (FoV) (e.g. $S_0 \sim
10^{-14}$ \fcgs for Sgr C).   In this way, we avoid the statistical
bias or fluctuation due to either low source statistics at the bright end
of the accessible flux range or limited sky coverage at the faint end
of the range. The portion of each
distribution not meeting the above criteria is dotted.

The slope ($\alpha$) of the \lnls distribution is calculated by a power law fit 
($N \propto S^{-\alpha}$)
to the solid line section of the distribution.  The $y$-axis error of
the distribution is given by the quadratic sum of the statistical error
(shown in the figure) and a constant systematic error ($\sim 20\%$,
the difference between the PL and PL+Fe model).  As expected for the narrow FoV
of ACIS-I observations, the slopes of the \lnls distributions
are largely consistent with the -1.5 slope within $\sim 2\sigma$ except
for the stacked Sgr A* field, which shows a hint of the actual deviation
($\sim 6\sigma$) from the -1.5 slope. Note the calculated slopes are
only for guiding purpose, and they should not be taken seriously for
representing the population since a simple power law does not fit some of
the distributions very well.

The AGN distribution is taken from \citet{Kim07}, using a power law
model with $\Gamma$ = 1.7 for the X-ray spectra.  The (orange) dashed
line indicates the reduced AGN population that can be seen through the
high extinction fields such as the Sgr A* field, since the unabsorbed flux is corrected for 
the average absorption of the X-ray sources, mostly Galactic and centered around the GC, 
which should be about half of the total absorption for the AGN. 
For simplicity, we correct another $\nH = 6 \nHu$ for the AGN seen in the high extinction
fields. 


For the Sgr A* field, we plot the results from both the stacked data
(black) and the 100 ks exposure (grey) in Fig.~5a and use only the
stacked data in Fig.~5b.  The spectral models from the 100 ks exposure
are used for  both data sets for fair comparison with other fields and
to avoid any spectral parameter driven variations between two exposures for
the Sgr A* field.  The distribution of the stacked data is $\sim 40\%$
higher  at $S_0 =
10^{-14} $ \fcgs than the same for the 100 ks
exposure.  A few factors such as the MB\footnote{Note that the unstacked 
Sgr A* field (Obs.~ID 3665) has the shortest GTI (88 ks) among the seven fields.} 
are responsible for the difference, but the main cause of the difference is suspected to be
the X-ray variability of the sources. The stacked data (750 ks) simply
have a better chance of detecting the sources or catching high flux
states of the sources than for the shorter exposure (100 ks).  For
instance, the 20 brightest sources in the \Hc band in the 100 ks
observation of the Sgr A* field are found to be about 30\% brighter on
average in the stacked data set, and five of the 20 brightest sources
in the stacked data were not detected in the 100 ks observation.  This
variation qualitatively agrees with the change seen in the \lnls
distributions, but the diverse nature of the X-ray variability and
duty cycles makes it hard to quantify the resulting difference in the
\lnls distributions.

The radial distribution is generated from the sources with $S> S_0 =
1.5 \times 10^{-14}$ \fcgs, plotted over the stellar distribution
(green) and the $1/\theta$ distribution (blue dashed).
Both the
stellar and $1/\theta$ distributions are averaged over the ACIS-I FoV
($17\arcmin \times 17\arcmin$) of the GB fields.
The $S_0$ value for the radial distribution is chosen as a compromise
between having sufficient source statistics in the Window fields ($S_0
\lesssim 2 \times 10^{-14}$ \fcgs) and avoiding statistical biases in
the high extinction fields ($S_0 \gtrsim 10^{-14}$ \fcgs).  The curve resulting
under the PL+Fe model (black solid) is more centrally concentrated around
the GC than the PL model (red dotted).

The stellar distribution is derived from the Galactic stellar models compiled
by M06, and it consists of a central spherical cluster ($\rho_A$, Eq.~1
in M06), a central disk ($\rho_B$, Eq.~2 in M06), a triaxial ellipsoidal
GB ($\rho_C$, Eq.~3 in M06) and a Galactic disk ($\rho_D$, Eq.~6 in M06;
for the origin of the formulae, see also Launhardt et al.~2002, hereafter L02;
Kent et al.~1991, hereafter K91).
For the first three components ($\rho_A$, $\rho_B$ \& $\rho_C$),
we use the formula and the parameter values in L02 and M06. For the
Galactic disk component ($\rho_D$), M06 use a simple exponential form in K91 and
employ 10\sS{11} \Ms for the total Galactic disk mass for the overall
normalization, but since the first three components are mainly for the
stellar mass, we believe this is an overestimate. Therefore, we use
a normalization that matches the local stellar mass density of 0.044 \Ms
pc\sS{-3} \citep{Robin03}. This gives $1.8 \times 10^{10}$ \Ms for
the whole disk, which is roughly consistent with the estimate by
\citet{Robin03} ($2.2 \times 10^{10}$ \Ms)\footnote{The small difference
is mainly due to the difference in the assumption of the distance to
the GC: 8 kpc for the model used here, and 8.5 kpc for
\citet{Robin03}.}.  Since we expect both the X-ray and stellar sources
are centrally concentrated around the GC, in Fig.~5b we further assume
that all detected hard Galactic  X-ray sources are at a distance of 6
-- 10 kpc, which is justified given that the stellar models predict
that $\gtrsim 80\%$ of sources along the line of the sight of the GB
fields lie in the same distance range.
Note the central concentration
also makes our normalization change of the Galactic disk component less
important in the outcome, but we find that the change makes this Galactic
stellar mass model consistent with other Galactic stellar number density
models (see \S\ref{s:density} and Table \ref{t:d}). These stellar model 
components have about a factor of two uncertainty (M06, L02).

The normalization of the stellar and $1/\theta$ distributions is set
by a simple $\chi^2$ fit to the radial-distribution curve under the
PL+Fe model (Fig.~5b). We use the stacked result for the Sgr A* field.
The radial distribution shows the GB X-ray sources are highly concentrated
at the GC, more than the stellar distribution.  It also shows the hard GB
X-ray sources extend out to $>$ 1.4\Deg from the GC, roughly following an
empirical relation of $1/\theta$ with some excess in the Arches Cluster,
Sgr C and Sgr B2 fields.

The excess of the X-ray source to stellar
distribution near the GC does not appear as prominent if we use the 100
ks exposure of the Sgr A* field at $S_0 =
1.5 \times 10^{-14}$ \fcgs. However, the trend of the relative excess
of the X-ray sources toward the GC is present from the Sgr B2 to Sgr A*
field (e.g. the Sgr B2 field has a deficit under the current relative
normalization in Fig.~5b), and the \lnls distributions of the 100
ks exposure and the stacked Sgr A* field become more consistent at
$S_0 \gtrsim 2 \times 10^{-14}$ \fcgs.  Therefore, the excess of X-ray
sources toward the GC with respect to the considered stellar model
appears real. We explore this excess in more detail in Section 5.1.

\section{Discussion} \label{s:disc}

We find that the number-density of the hard X-ray sources in the GB
is significantly elevated above the AGN density out to at least
the LW at 1.4\Deg  separation from Sgr A* (Fig.~\ref{f:lnls}a).
Furthermore, although empirical (see Fig.~\ref{f:model} for the
composition of the stellar population), the radial distribution of the
hard X-ray sources roughly follows a
$1/\theta$ relation out to this field (Fig.~\ref{f:lnls}b).  This
discovery suggests that all such sources observed within at least $\sim$
200 pc of the GC belong to the same centrally concentrated population.
The similarity of the stacked spectra of the hard X-ray sources in all
fields from Sgr A* to LW, and in particular the presence of a 6.7 keV
iron emission line, strengthens our conclusion that a single underlying
class of sources makes up this population.

\subsection{X-ray Source Density vs. CV density} \label{s:density}

The current leading candidate to explain the X-ray sources within 20 pc
around the GC is magnetic CVs or intermediate polars (IPs) in particular (M04, L05).
Recent population synthesis models by \citet[hereafter R06]{Ruiter06} show IPs can
constitute the majority of these X-ray sources under the assumption that
IPs span a luminosity range of $\sim 3 \times 10^{29} - 5 \times
10^{33}$ \lcgs and that they make up $\sim 2 -8\%$ of all CVs (see also
M06 for a review of the population synthesis models for the X-ray sources
in the GB).  Now our radial distribution indicates 
this source population extends out to $\sim
200$ pc and the hard X-ray spectra with the iron emission line supports
the idea that IPs are the major component of the population.

In this section, we compare the observed X-ray source density with
the stellar (mass) density to see if CVs, especially  IPs can
explain the majority of the detected X-ray sources.  
Table \ref{t:d} summarizes the number density of the X-ray sources with
$S> 1.5 \times 10^{-14}$ \fcgs in the \Hc band and compares them with the
stellar (mass) density.  For the GB X-ray source densities, we subtract
the expected number of AGN, which is 145 deg\sS{-2} in the Window fields
and 107 deg\sS{-2} in the high extinction fields, from the surface density
(see \S\ref{s:lnls}) \citep{Kim07}.  

Table \ref{t:d} quotes the average stellar density over the volume defined
by the distance between 6 and 10 kpc in the $17\arcmin \times 17\arcmin$
FoV using two stellar models.  Stellar model A is the same stellar mass
model used in Fig.~\ref{f:lnls} (M06, L02, K91).  Since model A
provides the stellar
mass density, we use a local value of 0.144 stars pc\sS{-3} and 0.044
\Ms pc\sS{-3} to convert the mass density to the number density or
vice versa \citep{Picaud04,Robin03}.  As a consistency check, we also compute
the stellar density using another stellar model (model B) by
\citet{Picaud04}, which consists of a Galactic
disk and a Galactic Bulge. This model describes the stellar number
density
in the outer Galactic Bulge and the Galactic disk. So it is
properly normalized at the local Galaxy (0.144 star pc\sS{-3}),
but due to the lack of the Galactic nucleus components ($\rho_A \&
\rho_B$), the stellar density for the GB fields within 1\Deg of the GC
is underestimated.  The two models agree within 30\% for the Window
fields where Galactic nucleus components are relatively unimportant.
In the following, we use model A for comparing the X-ray source density
with the stellar density.

The relative X-ray source to stellar mass densities in the seven GB fields
are $0.3 - 1.8\times 10^{-6}$ X-ray sources \Msi at $S>1.5\times
10^{-14}$ \fcgs ($1.1 \times 10^{32}$ \lcgs for sources at the GC, 8 kpc).
The large variation of the relative density among the seven fields
reflects the mismatch between the X-ray and stellar distributions - the
X-ray sources are more centrally concentrated than the stellar sources.
The relative X-ray source to stellar number densities
are $0.9-5.7 \times 10^{-7}$ X-ray sources star\sS{-1} at $S>1.5\times
10^{-14}$ \fcgs, depending on the field.  

Now we assume IPs are 5\% of all CVs (e.g. $\sim$ 2--8\% for the
models in R06) and about 12\% of IPs have the X-ray luminosity above
$10^{32}$ \lcgs (e.g.  $\sim$ 10--16\% in R06, see also
\citet{Heinke08}).  Then the required CV to stellar density to explain
the hard X-ray GB sources ranges from 1.6 to 9.5 $\times 10^{-5}$
depending on the fields.  If we assume a local star density of 0.144
pc\sS{-3}, these
correspond to the equivalent local CV density of $0.2-1.4 \times 10^{-5}$
pc\sS{-3}. Considering the current local CV density estimates ($1-3 \times
10^{-5}$ pc\sS{-3}) in the literature
\citep[see e.g.][]{Ak08,Grindlay05,Pretorius07}, this result indicates
IPs can be the major component of the observed X-ray sources as long
as the relative CV to stellar density in the GB is comparable to the
value in the local solar neighborhood. 

Note there are a few caveats in this analysis.  First, the radial
distribution of the X-ray sources does not match well with the stellar
distribution as shown in Fig.~5b.  As mentioned, this is the reason
for the large variation in the estimates of the relative CV density.
It means the stellar model we use may not be appropriate for scaling the
observed X-ray population directly. A solution could be found in some of
the assumptions we have made. For instance, the (apparent) fraction of
IPs in all CVs may not be constant across the fields.

Second, there are large uncertainties in the model parameters and various
assumptions such as the ratio of IPs to all CVs and
the fractional IPs with the X-ray luminosity $\ge 10^{32}$ \lcgs.
For instance, according to \citet{Ritter03}, the ratio of the known IPs
to all known CVs are about 10\%, but this is also subject to a large
uncertainty due to selection biases.  
Similarly there is no firm estimate of the X-ray IP
luminosity distribution to set the accurate limit for the fractional
IPs with the X-ray luminosity $\ge 10^{32}$ \lcgs.

Third, as illustrated in the \lnls distributions of the
100 ks and stacked data of the Sgr A* field, the X-ray variability can
change the apparent source distribution. In order to understand the true
distribution, it is necessary to monitor the GB fields continuously and
extract the source distribution from a longer exposure.
Considering the X-ray variability of the sources observed in the Sgr A*
field, the true distribution of the GB X-ray population in the other GB
fields can be $\gtrsim$ 20--30\% higher than what has been observed in
the 100 ks exposures.

\begin{figure*}[t] 
\plotone{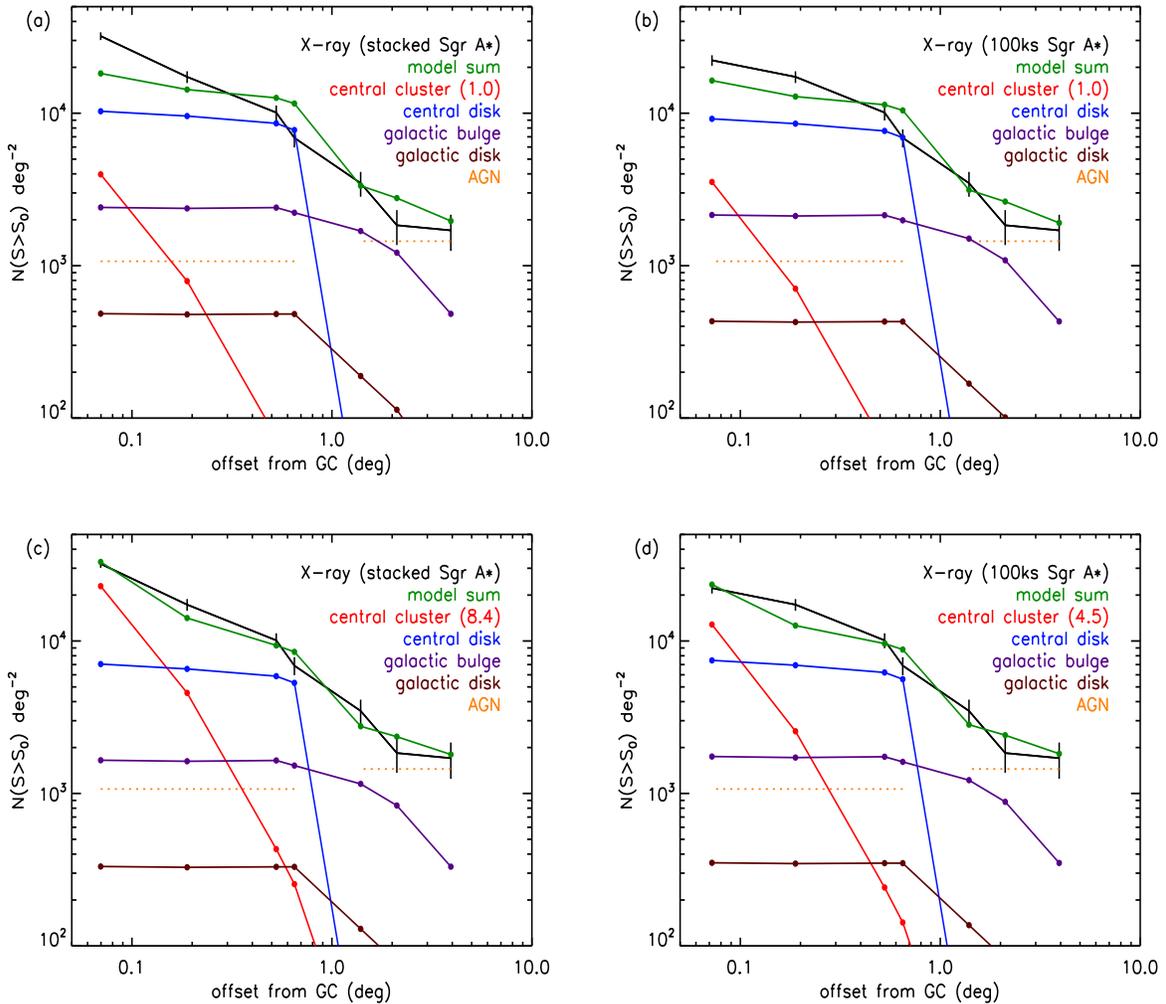}
\caption{Comparison of the X-ray and stellar distributions: the best model fit
of the stellar to X-ray distrubtion with (a) the stacked Sgr A* data, 
and (b) the 100 ks Sgr A* data, and (c \& d)
the same model fit but with the freed normalization parameter of the
central spherical cluster component ($\rho_A$). The excess of the central spherical cluster
component ($\rho_A$) needed for the best model fit is 8.4 or
4.5$\times$ higher than the original model, depending on the Sgr A* data set.
}
\label{f:model} \end{figure*}

A more recent
study by \citet{Schodel07}  of the stellar population and mass content in
the Galactic nucleus  shows evidence for a higher mass contained in the central
$\sim$ 1 pc than predicted by studies and they speculate the excess may
be due to a stellar remnant population such as black holes.
Fig.~\ref{f:model} shows the composition of the stellar components
($\rho_A$, $\rho_B$, $\rho_C$ \&
$\rho_D$) in comparison with the observed projected X-ray source density.
Fig.~\ref{f:model}a shows the fit results using the stacked Sgr A* data
(the same as Fig.~\ref{f:lnls}b)
and Fig.~\ref{f:model}b for the 100 ks Sgr A* data.  In order to allow the
possible excess of the X-ray population in the Galactic nucelus relative
to the observable stellar population, we also fit the X-ray distribution
by freeing the relative normalization parameter of the central spherical
cluster component ($\rho_A$) with respect to the rest of the components
(c \& d in Fig.~\ref{f:model}).  The resulting fits are substantially
better, but they require a large excess of the central stellar cluster
component relative to the fixed normalization given in Section 4.3:
8.4$\times$
more for the stacked Sgr A* data and 4.5$\times$ more for the 100 ks
observation.  This excess is still larger than the mass excess ($\sim$
2$\times$) of the nucleus above the observable stellar cluster estimated
by \citet{Schodel07}, although the latter has a large uncertainty ($\sim
30 - 50$\%).

\subsection{Total Surface Brightness}\label{s:surface}

\begin{figure*}[t] 
\plotone{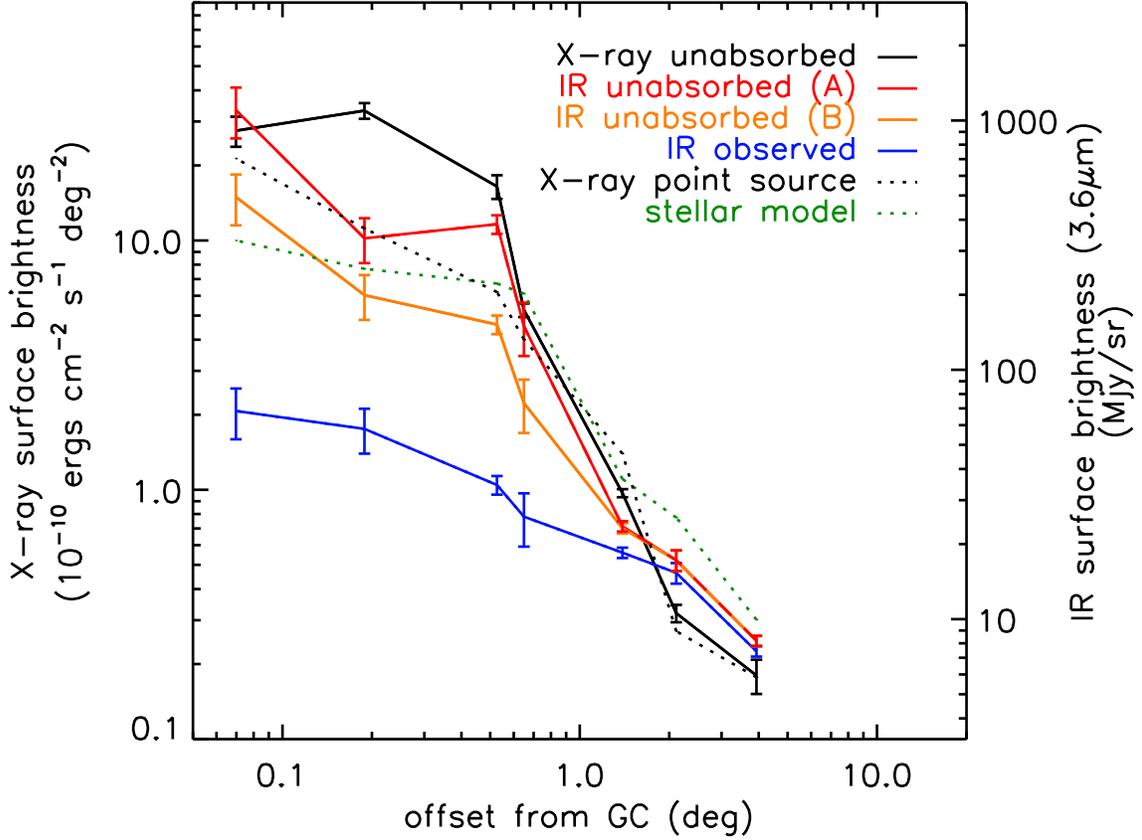}
\caption{Comparison of the X-ray and the IR surface brightness.  The X-ray
surface brightness (black solid line) is the total unabsorbed \Hc flux minus
the expected CXB flux in the \chandra ACIS-I FoV of each field
assuming the quantile average
spectral model (PL+Fe) of the \Gm (for the Windows fields) or \Gh
group (for the rest). The infrared surface brightness (blue solid) is
the average flux in the \chandra ACIS-I
FoV from the 3.6 $\mu$m band images by \spitzer/GLIMPSE. To calculate
the unabsorded IR flux consistently with the X-ray flux,
we assumed the \nH estimates by the quantile analysis in Table \ref{t:q}
(\Gm for the Windows fields and \Gh for the rest). We use the \nH
estimates by both models (A for PL and B for PL+Fe). We also assumed
$A_{3.6\mu \mbox{\scriptsize m}} = 0.5\  A_{Ks}$ \citep{Nishiyama09}.
The errors of the X-ray brightness are the quadratic sum
of the statistical and systematic errors and the errors of the CXB
estimates (see \S\ref{s:surface}).
The errors of the IR brightness are based on the statistical
fluctuation of the brightness in the four ACIS-I fields, but the dorminant
errors in the high extinction fields come from the uncertainty in the
extinction estimates as seen in the large difference of lines A and B.
For easy comparison, we overlay the distributions of the X-ray source
density (black dotted) and what the stellar model predicts (green
dotted) without the AGN component using an arbitary normalization factor.
}
\label{f:surface} \end{figure*}

Since the IR surface brightness is a good indicator of the stellar
population, a direct comparison of the IR and X-ray surface brightness
provides an independent clue on the origin of the X-ray emission in
the fields without influence of the uncertainties in the stellar model.
In Table~\ref{t:surface} we list the total
X-ray and IR surface brightness of the seven fields. In
Fig.~\ref{f:surface} we compare the total IR surface brightness with
the X-ray surface brightness after removing the contribution from the
cosmic X-ray background (CXB). The CXB is expected to be $(1.6-1.7
\pm 0.2) \times 10^{-11}$  \fcgs deg\sS{-2} for the Windows fields and
$(1.3-1.4 \pm 0.2) \times 10^{-11}$ \fcgs deg\sS{-2} for the rest \citep{Hickox06}.
For easy comparison,
Fig.~\ref{f:surface} also overlays the distributions of the X-ray source
density and what the stellar model predicts without the AGN component
in the dotted lines.  

The X-ray surface brightness is calculated from
the integrated counts in the \Hc band, using the quantile
averaged model spectra of the \Gm group for the Windows fields and the
\Gh group for the rest (Table~\ref{t:q}). In order to
subtract the instrumental background, we use the CXC background database
constructed from stowed observations.  For the relative normalization of
the background subtraction, we use the the integrated counts in the 10.5
- 12.5 keV.  The error bars in Fig.~\ref{f:surface} are the quadratic
sum of the statistical and systematic errors and the errors of the CXB
estimates. For the statistical errors, we use the difference in the flux
estimates between the PL and PL+Fe models.  

The IR surface brightness
is calculated from the 3.6 $\mu$m mosaic images (1.2\arcsec pixel)
generated by Galactic Legacy Infrared Mid-Plane Survey Extraordinaire
(GLIMPSE) from the \spitzer/IRAC observations
\citep{Churchwell09}. The error is based on the statistical variation
of the surface brightness in the four \chandra ACIS-I FoVs of each
field.  In order to estimate the unabsorbed surface brightness
consistently with the X-ray flux estimates,
we assume the \nH estimates by the quantile analysis in Table \ref{t:q}
(\Gm for the Windows fields and \Gh for the rest). We use the \nH
estimates by both models (line A for PL and line B for PL+Fe in
Fig.~\ref{f:surface}) to explore the dependence of the \nH estimates.
For a given \nH estimate,
we use $\nH = 1.79 \times 10^{21} A_V$, $A_K = 0.114 A_V = 0.95 A_{Ks}$,
and $A_{3.6\mu \mbox{\scriptsize m}} = 0.5\ A_{Ks}$ \citep{Nishiyama09}.

The radial distribution of the CXB-subtracted X-ray surface brightness
roughly matches with the distribution of the unabsorbed IR surface
brightness under the given uncertainties with the possible exception
of the Arches field.  This indirectly supports the idea that they share
a common origin.  In particular, the distribution of the unabsorbed IR
surface brightness matches with the X-ray source distribution, showing
higher concentration at the GC than the stellar distribution model.
The origin of the discrepancy between the stellar distribution model and
the IR surface brightness is unclear and requires further investigations,
but note that the unabsorbed IR brightness in the high extinction
fields depends sensitively on the extinction estimates, resulting in
relatively large uncertainties.  This dependence is also evident in
the large variation among the ACIS-I FoVs in each field, which is at
least in part due to the variation in the extinction across the FoVs.
The radial distributions of the total X-ray surface
brightness (Fig.~\ref{f:surface}) and the X-ray source density
(Fig.~\ref{f:lnls}) also shows a slightly different trend in a few
fields such as the Arches field. For instance, the difference in the
Arches field is due to the large contribution of a few bright sources
in the total flux.  We will address the detail of the unresolved X-ray
emission with respect to the identified point sources
in a following paper \citep{Hong09b}.



\subsection{Comparison with other results} 

According to Eq.~5 in M03, the X-ray source density in the Sgr A* field
is 0.60 $\pm$ 0.04 X-ray sources arcmin\sS{-2} at $S> 1.5 \times
10^{-14}$ \fcgs or $1.25 \times 10^{-7}$ \pcgs in the 2--8 keV range
under their assumption of a power law spectrum with $\Gamma=0.5$ and $\nH=6\times 10^{22}$
cm\sS{-2} \footnote{M06 assume $\Gamma=1.5$ for the X-ray spectra of the 
sources in the $2\Deg \times 1\Deg$ region around the GC}.
This is roughly consistent with our results, 0.86  $\pm$ 0.17
arcmin\sS{-2} at $S > 1.5\times 10^{-14}$ \fcgs from the stacked results
under the PL+Fe model. The error is derived from the quadratic sum of
the statistical error and $\sim$ 20\% systematic errors (the difference
between the PL and PL+Fe model).

Table \ref{t:sl} summarizes a few estimates of the specific luminosity of
the Galactic X-ray point sources in the \chandra/ACIS energy range in the
literature.   The range of the specific luminosity in our study is
the variation among the seven
fields under the assumption of the PL+Fe model for the X-ray spectra.
Using Eq.~7 in M06 and assuming the $\alpha$ values in Fig.~5a and the
number density of the X-ray sources ($0.3-1.8\times 10^{-6}$
X-ray sources \Msi at $> 1.1 \times 10^{32}$ \lcgs) in Table 4, we get
$0.5 - 2.8 \times 10^{26}$ \lcgs  \Msi in $2-8$ keV for the luminosity
range of $5\times 10^{32} - 10^{34}$ \lcgs.  

M06 interpreted the result in \citet[hereafter S06]{Sazonov06} to be $1.0\pm 0.3 \times
10^{27}$ \lcgs \Msi for $10^{32.7-34}$ \lcgs using Eq.~5 in S06 and
Eq.~7 in M06, and claimed their result ($5 \pm 2 \times 10^{27}$
\lcgs \Msi) is consistent with S06.  However, the result in S06 is
calculated in the 2--10 keV range and M06 in 0.5--8 keV.  In the $2-8$
keV range, the result in M06 becomes $3.3 \pm 1.3 \times 10^{26}$
\lcgs \Msi under their assumption of $\Gamma=1.5$ and $\nH=6 \nHu$.
Similarly, the result in S06 is scaled to be $9\pm 3 \times 10^{26}$
\lcgs \Msi in the same energy band. So there is a hint of mismatch in
the results between M06 and S06 if one follows the scaling in M06, but
also note Fig.~9 in S06 shows $\sim 6 \pm 2 \times 10^{26}$ \lcgs \Msi in
the 2--10 keV band for $>10^{32-33}$ \lcgs, which is consistent with M06 and
lower than the scaling done for S06 in M06.
This conversion also reveals the
result in M06 is consistent with our result for the Sgr A* field. 
Using a similar scaling based on Eq.~5 in S06, we get $3 \pm 1 \times
10^{26}$ \lcgs \Msi for the X-ray emissivity reported by
\citet[][hereafter R07b]{Revnivtsev07b} in the $2-8$ keV range for $10^{32.7-34}$ \lcgs, assuming
3\% of tht total emission in the same luminosity range based on Fig.~6 in R07b. 
Due to many different underlying assumptions in the above estimates
(e.g. the spectral model parameters, stellar models, etc), it is not
easy to make a fair comparison among the reported results.   The large
uncertainties make these results appear consistent within 2$\sigma$,
but our results are at the lower end of these findings.

M06 and \citet[hereafter M08]{Muno08} have presented the X-ray source
distribution in a $2\Deg \times 1\Deg$ region around the GC.  
In the case of the \lnls distribution, one of the interesting results in
M06 and M08 is  a flatter distribution of the X-ray sources in the Arches
Cluster and the subsequent excess of the X-ray sources near the high end
of the flux range, compared to the Sgr A* field.  We also see
a similar cross over between the unstacked Sgr A* field and the Arches
Cluster at $\sim 1.5 \times 10^{-13}$ \fcgs (or at $\sim 8 \times
10^{-13}$ \fcgs with the stacked Sgr A field, out of the range in
the Fig.~5a).  We believe this is a simple statistical fluctuation rather
than a true representative of the population, arising from
a small number of sources in the narrow FoV,
where a few strong sources ($\sim 2 - 4$ in the Arches Cluster) skew
the shape of the whole distribution.  In fact, the small number statistics is also evident in the jumpy shape
of the distributions near the high end of the flux range. In Fig.~5 of M06, the excess of the X-ray
sources in the Arches Cluster above $6 \times
10^{-6}$ \pcgs is boosted by excluding the
overlapping region between the Sgr A* field and the Arches Cluster from
the sky coverage calculation for the X-ray sources of the Arches
Cluster. In order to minimize
the effects due to the small number statistics, in our analysis, we
focus on the solid line  section of the distributions that contain
at least 10 or more sources.  In addition, quantile analysis results
in a slightly higher value ($\sim 10$\%)
of the rate-to-flux conversion factor for the \Gh sources in the Sgr A*
field than the same for the Arches Cluster (see Fig.~4), which in turn pushes 
the \lnls distribution of the Sgr A* field relatively higher.
As a result, we find the slopes of the \lnls distributions of the 
Arches Cluster and Sgr A* fields are consistent and the Sgr A* field
contains more X-ray sources than the Arches Cluster consistently below
$\sim 3 \times 10^{-14}$ \fcgs.

In the case of the radial distribution, the excess of the X-ray sources
in the Sgr A* field with respect to the stellar model is about 3$\sigma$
above the stellar model if one considers both the statistical errors
and the $\sim 20\%$ systematic errors in the flux estimates (about
10$\sigma$ above only with the statistical errors).  M08 find about
2.5$\sigma$ excess of the X-ray sources at the GC compared to the best
fit stellar model.

\subsection{Another source population in Sgr C (or Sgr B2)?} \label{s:sgrc}

The $1/\theta$ distribution is roughly consistent with the observed radial
distribution of the X-ray sources in the GB fields within 2--3$\sigma$
except for the Sgr C field.   The $1/\theta$ distribution is a merely
empirical outcome if the observed X-ray source density consists
of multiple components.  On the other hand, the apparent $1/\theta$
distribution along with the similar X-ray spectral properties may imply
that the GB X-ray population belongs to a homegenous component. If true,
the excess of the Sgr C and Sgr B2 fields can be simply viewed as the
presence of another source population in these fields in addition to
the population following the $1/\theta$ distribution.  The Sgr C field,
like the Sgr B2 field, contains molecular HII complexes that host massive
star formation. These molecular clouds are very luminous in hard X-rays,
in particular with the 6.4 keV neutral iron line
\citep{Murakami01a,Murakami01b}.  In our analysis, the estimates of the
$\Gamma$ value in the PL and PL+Fe model for the stacked spectra of the
\Gh sources in the Sgr B2 and C fields, are relatively lower compared
to  the rest of the fields, suggesting the possibility of another source
population with a different spectral type that could be related to the
star formation.

\section{Conclusion and Future Work}

In the \lnls distribution of the sources in the GB fields, the systematic
errors arising from certain assumptions of spectral type is usually
disregarded due to lack of alternative approaches. However they often dominate
other systematic errors such as the EB, completeness or even statistical
errors.  The quantile analysis allows for a simple, robust method to assign
a proper spectral type for flux calculation. The technique is shown to
be reliable in the hard band ($>2$ keV) and insensitive to selection of the
spectral model.  In the soft band ($<2$ keV), where the Galactic
extinction has a great influence in the spectra, the result can vary
drastically depending on the assumed spectral model class.  Therefore,
any results covering the soft energy range should be taken with caution.

The \lnls and radial distribution of the GB fields including the three low
extinction Windows show the high concentration of the GB X-ray sources
near the GC.  The GB distribution clearly extends out to $\sim$ 1.4\Deg
(LW) from the GC and possibly more.  The spectral type of the GB X-ray
sources appears to be largely consistent across the region under the power
law model with an iron emission line at 6.7 keV. It is possible that one
type of source constitutes the majority of the GB population, and the
estimated X-ray density is consistent with the majority being magnetic CVs
(IPs). The $1/\theta$ relation of the radial distribution of the X-ray
sources appears to be simply empirical in comparison with the stellar
model compositions. Since the gravitational influence of the central
supermassive black hole is only dominant within a few central pc (a few
arcmin) at most, there is no compelling physics behind the $1/\theta$
relation being fundamental out to a few degrees from the GC.  However,
the discrepancy between the stellar models and the X-ray distribution,
and the apparent homogeneity of the X-ray spectral properties of these
X-ray sources pose other possibilities: the $1/\theta$ relation may not
be so empirical or at least the X-ray source population contains some
components that are not easily traceable by the visible stellar population.

The radial distributions of the total X-ray and IR surface brightness
in the fields match within the given uncertainties, implying the same
origin. The radial distribution of the IR surface brightness resembles the
distribution of the X-ray point sources perhaps better than predicted by
stellar distribution models, but a further detailed analysis is required
because of the sensitive dependence of the IR surface brightness on
the extinction estimates in the high extinction fields. In the case of
the X-ray surface brightness, the additional care must be taken due to
dominant contributions of a few bright sources on the X-ray flux in star
formation fields such as The Arches field.

While multiwavelength observational campaigns provide important clues
on the GB X-ray population, the true nature of the nature of GB X-ray
sources may not be completely  resolved due to source confusion and
high obscuration.
A deep observation ($\sim 1 $ Ms) of the LW \citep{Revnivtsev09}, designed
to investigate the nature of the Galactic Ridge X-ray emission (GRXE) in
the field, is very encouraging for studies of the nature of X-ray point
sources in the GB.  Such a deep observation allows a direct detection
of iron emission lines or X-ray variability in many of the GB X-ray
sources, with which we can identify the nature of individual sources.
We note only a handful of sources in the 1 Ms data of the Sgr A* field
had enough statistics for identification through such a direct discovery
(M03,M04).  But the low extinction in the Window fields can be a game
changer.  For instance, we have identified an IP in the BW from the 100
ks observation, based on the periodic X-ray modulation associated with
the X-ray spectral
change \citep{Hong09a}.  According to its average flux, the source can be
a bright IP ($\sim$ 10\sS{33} \lcgs) near the GC, but for a similar source
in the Sgr A* field it would be very difficult to observe such a spectral
change or periodic modulation due to the heavy absorption.  By a crude
scaling based on one IP found in the 100 ks observation of the BW, one can
expect about 30 to 40 such identifications in a 1 Ms exposure of the
BW\footnote{Assume the identifiable source distribution is proportional to $S\Ss{th}\sS{-3/2}$,
where $S\Ss{th}$ gets 10 times fainter, and also assume an additional
20-30\% increase in the probability of catching highly variable X-ray
sources based on the difference in the stacked and unstacked data set of the Sgr A* field.}. 
Such findings would also provide enough statistics to explore the radial 
distribution of this particular source type.  Therefore, continuous X-ray monitoring of the low
extinction Window fields including the SW and BW is another important
approach for unveiling the nature of the GB X-ray sources.  Note that
the Window fields are also suitable for searching non-magnetic CVs in
the GB.  These are potentially more abundant than magnetic CVs,
but they are known to have relatively soft spectra and thus they would
be likely hidden in the high extinction fields such as the Sgr A* field.


This work is supported in part by NASA/Chandra grants GO6-7088X, GO7-8090X
and GO8-9093X.  We thank the referee for the insightful comments and
discussions.

\begin{table}
\small
\caption{X-ray point sources in the selected GB fields} 
\begin{tabular*}{\textwidth}{l@{\extracolsep{\fill}}lrrrrrrrrrrrrr}
\hline\hline
Field  	& Obs. ID	&$l$ 		&$b$		& \sS{a}Offset	& \sS{b}\nHt 	&\sS{c}GTI 	& \multicolumn{4}{c}{Source count\sS{d}} 	\\	
\cline{8-11}
	& 		&(\Deg)		&(\Deg) 	& (\Deg)	&		&(ks)		& \Bx	&\Sx	& \Hx	& Combined		\\
\hline
BW	&3780		&1.06		&$-$3.83 	& 3.93		&0.31 		&96 		&  365	& 326	&  134	& 407   		\\			
SW	&4547, 5303	&0.25 		&$-$2.15 	& 2.12		&0.48 		&96 		&  388	& 313 	&  140	& 433   		\\
LW	&5934, 6362,6365&0.10 		&$-$1.43 	& 1.39		&0.68 		&94 		&  282	& 184 	&  174	& 319   		\\
\hline
Sgr B2	&944		&0.59 		&$-$0.03 	& 0.65		&81.2 		&97 		&  279	& 126 	&  224	& 363   		\\
Sgr C	&5892		&$-$0.57 	&$-$0.02 	& 0.51		&52.7 		&97 		&  313	& 188 	&  241	& 442   		\\
Arches  &4500		&0.12 		&$-$0.02	& 0.19		&52.5 		&97 		&  330	& 84 	&  328	& 423   		\\ \hline
Sgr A*	&3665		&$-$0.06 	&$-$0.05 	& $-$		&56.5 		&88 		&  401  & 92	&  400	& 508   		\\ 
(Stacked\sS{e})& 	&		& 		& 		&		&698		&  2251 & 370	&  2316	&2876  		\\ \hline
\end{tabular*}
(a) The aim point offset from Sgr A*.
(b) The estimates for the integrated neutral hydrogen column density along
	the line of the sight (in $10^{22}$cm$^{-2}$) by \citet{Schlegel98} for the location of the aimpoint. 
	This is only for guiding purpose due to the large uncertainty in the Galactic plane fields.
(c) The good time intervals (GTIs).  The total exposure (i.e. before
	cleaning) is 100 ks each (750 ks for the stacked Sgr A* field).
(d) The number of the sources with net count $\ge 1$ in the broad band
(0.3--8 keV) on the ACIS-I CCDs (0, 1, 2, and 3) in the three
detection bands (\Bx: 0.3--8 keV, \Sx: 0.3--2.5 keV, \Hx: 2.5--8 keV)
and the combined unique source list. 
(e) 14 pointings are stacked, and they are Obs.~ID 242, 2951, 2952, 2953, 2954, 2943, 3663, 3392, 3393, 3549, 3665, 4683, 4684
and 5360.
\label{t:f}
\end{table}
\clearpage

\begin{sidewaystable*}
\scriptsize
\caption{Catalog of X-ray point sources in the Window and four GB fields} 
\begin{tabular*}{\textwidth}{l@{\extracolsep{\fill}}rrrrrrrrrrrrrrrr}
\hline\hline
Source	  	& 	&  		&		& Posi.		&  		& \multicolumn{3}{c}{Net counts\sS{c}}	&$S/N$\sS{d}	& \multicolumn{3}{c}{Quantiles} 	& {Unabsorbed Flux\sS{f}}	\\
\cline{7-9}\cline{11-13}
Name 	  	& Field	& R.A. 		&Dec.		& Error\sS{a}	& Offset\sS{b} 	& \Bx	&\Sc		& \Hc	 	&  \Hc		& $E_{50}$ 	& Quartile 	& Group	& \Hc					\\	
(CXOPS J) 	&  	& (\Deg)	&(\Deg) 	& (\arcsec)	& (\arcmin)	& 	& 		&		&     		& (keV)		& Ratio\sS{e}	& 	&{($10^{-14}$ \fcgs)}	\\
\hline
180230.4-295647 &BW &  270.626934 &  -29.946497 &  1.29 & 10.05 &  123.6 (14.3) &   71.5 (10.9) &   53.3 (9.9) &     5.4 &   1.89 (0.10) &   1.30 (0.17) &  1 &   1.31 (0.24) \\
180231.2-295528 &BW &  270.630007 &  -29.924698 &  2.37 & 10.12 &   38.6 (10.9) &    4.6 (6.9) &   34.8 (8.9) &     3.9 &   3.56 (0.29) & 1.68 (0.30) &  3 &   1.12 (0.29) \\
180235.9-295323 &BW &  270.649946 &  -29.889846 &  3.32 &  9.87 &   23.6 (9.6) &   23.4 (8.1) &   -0.2 (5.6) &    0.0 &   1.05 (0.15) & 2.08 (0.48) &  1 &  -0.01 (0.13) \\
 .... \\
175404.4-294359 &SW &  268.518385 &  -29.733089 &  2.56 &  9.58 &   34.1 (11.2) &   25.3 (9.2) &    5.9 (6.6) &     0.9 &   1.34 (0.18) & 1.63 (0.64) &  1 &   0.13 (0.14) \\
175405.3-294717 &SW &  268.522117 &  -29.788307 &  2.47 &  8.04 &   22.2 (8.9) &   19.7 (7.7) &    2.9 (5.1) &     0.6 &   1.40 (0.24) & 1.61 (0.40) &  1 &   0.06 (0.11) \\
175406.7-294239 &SW &  268.527957 &  -29.711050 &  1.97 &  9.99 &   42.2 (12.0) &   20.6 (9.4) &   22.5 (8.0) &     2.8 &   2.12 (0.54) & 1.08 (0.32) &  2 &   0.69 (0.24) \\
 .... \\
175051.2-293418 &LW &  267.713518 &  -29.571797 &  1.02 &  8.10 &  113.8 (13.8) &   36.2 (9.1) &   75.8 (10.9) &     7.0 &   2.63 (0.20) &   1.25 (0.16) &  2 &   2.21 (0.32) \\
175052.0-293319 &LW &  267.716827 &  -29.555400 &  2.92 &  8.14 &   16.4 (9.2) &    5.2 (6.8) &    9.3 (6.6) &     1.4 &   0.97 (5.31) & 0.20 (0.29) &  2 &   0.27 (0.19) \\
175053.3-293207 &LW &  267.722097 &  -29.535548 &  2.12 &  8.29 &   25.0 (10.3) &    2.8 (7.3) &   22.4 (7.8) &     2.9 &   3.46 (0.38) & 1.74 (0.48) &  3 &   0.77 (0.27) \\
 .... \\
\hline
\end{tabular*} Notes.---This table shows a part of the complete
list, which is available in the electronic edition. \\ 
(a) The 95\% positional error radius. 
(b) The offset from the aim point. 
(c) The net counts based on the aperture photometry\citep{Hong05}. 
(d) The signal to noise ratio in the \Hc band. The sources with $S/N\ge3$ are
	included in the \lnls plot in Fig.~5.  
(e) $3 (E_{25} - 0.3\ \mbox{keV})/(E_{75} - 0.3\ \mbox{keV}).$ 
(f) Based on the PL+FE model using quantile analysis.  We do
not include the flux estimates in the other bands due to their large
uncertainty. See the text for the details.  
The uncertainties for net counts and fluxes are statistical errors.
\label{t:sc}
\end{sidewaystable*}
\clearpage

\begin{table*}
\scriptsize
\caption{spectral model parameters for the \Gm and \Gh sources} 
\begin{tabular*}{\textwidth}{l@{\extracolsep{\fill}}cccccccccc}
\hline\hline
	& 		& Quartile  	& \multicolumn{2}{c}{PL from }			& \multicolumn{2}{c}{PL+Fe (He-$\alpha$) from}	&\multicolumn{4}{c}{PL (+Fe He-$\alpha$) from}	\\
Field  	& 		& Ratio\sS{a} 	& \multicolumn{2}{c}{Quantile Diagram\sS{b}}	& \multicolumn{2}{c}{Quantile Diagram\sS{c}}	&\multicolumn{4}{c}{Spectral Model Fit\sS{d}}	\\
	& $E_{50}$ 	&	      		& $\Gamma$   		& \nHt		& $\Gamma$   		& \nHt		& $\Gamma$    		& \nHt		& EW\sS{e}	& $\chi^2$/DoF\sS{f} \\
	& (keV)	 	&	      		&    			& (\nHu)	&    			& (\nHu)	&     			& (\nHu)	& (keV)		& 	\\
\hline
\multicolumn{9}{l}{unabsorbed hard sources (\Gm)} \\
BW 	&    1.97(2)    & 1.02(1)     		& {\bf 1.38}(03)       & 0.31(03)	& {\bf 1.42}(03)       & 0.32(05)	& {\bf 1.36}(2)       & 0.26(01)     	& 0.15(9)	& 113.3/167\\
SW 	&    2.10(5)    & 1.03(2)      		& {\bf 1.35}(07)       & 0.37(06)	& {\bf 1.38}(10)       & 0.38(08)	& {\bf 1.22}(3)       & 0.25(03)      	& $-$\sS{g}	& 48.7/69\\
LW 	&    2.66(4)    & 1.09(2)    		& {\bf 1.28}(07)       & 0.73(09)	& {\bf 1.35}(07)       & 0.79(09)	& {\bf 0.99}(2)       & 0.38(02)      	& $-$\sS{g}	& 125.9/148\\
\hline
\multicolumn{9}{l}{absorbed hard sources (\Gh)} \\
BW 	&    3.22(13)   & 1.38(9)      		& {\bf 1.66}(37)       & 2.20(70)	& {\bf 1.74}(37)       & 2.30(70)	& {\bf 1.22}(4)       & 1.66(16)     	& $-$\sS{g}	& 24.8/26\\
SW 	&    3.39(7)    & 1.52(4)    		& {\bf 1.77}(23)       & 2.90(50)	& {\bf 1.91}(23)       & 3.10(50)	& {\bf 1.58}(4)       & 2.78(19)      	& $-$\sS{g}	& 35.4/22\\
LW 	&    3.48(5)    & 1.43(3)      		& {\bf 1.21}(10)       & 1.95(15)	& {\bf 1.32}(10)       & 2.10(20)	& {\bf 1.30}(2)       & 1.89(06)      	& 0.17(8)	& 93.1/115\\
Sgr B2 	&    4.75(4)    & 1.86(2)      		& {\bf $-$0.37}(14)    & 3.40(80)	& {\bf 0.25}(17)       & 5.70(90)	& {\bf 0.50}(1)       & 6.20(22)     	& 0.61(7)	& 105.2/154\\
Sgr C 	&    4.81(3)    & 1.90(2)      		& {\bf $-$0.26}(10)    & 4.8(1.0)	& {\bf 0.46}(21)       & 7.8(1.2)	& {\bf $-$0.10}(1)    & 3.95(18)      	& 0.38(5)	& 172.3/189\\
Arches 	&    4.54(2)    & 1.83(1)      		& {\bf 0.14}(07)       & 4.00(50)	& {\bf 0.67}(14)       & 5.70(70)	& {\bf 0.85}(1)       & 5.17(12)      	& 0.66(5)	& 319.4/363\\
Sgr A* 	&    4.69(2)    & 1.91(1)      		& {\bf 0.31}(14)       & 6.40(60)	& {\bf 0.94}(14)       & 9.00(80)	& {\bf 1.02}(1)       & 6.95(12)      	& 0.46(4)	& 324.3/364\\
\hline
\end{tabular*}
(a) $3 (E_{25} - 0.3\ \mbox{keV})/(E_{75} - 0.3\ \mbox{keV}).$ 
(b) The parameter estimates based on quantile analysis for a power law model.
(c) The same as (b) but with a fixed 0.4 keV EW at 6.7 keV 
(d) The parameter estimates by the spectral model fit. (e) the EW of 6.7 keV line. 
(f) Degrees of Freedom.
(g) Due to poor statistics, the spectral fit is done with a power law model without an iron line. See \S\ref{s:dist}.
\label{t:q}
\end{table*}
\clearpage

\begin{table*}
\scriptsize
\caption{X-ray source and stellar density\sS{a}}
\begin{tabular*}{0.98\textwidth}{l@{\extracolsep{\fill}}cccccccccccc}
\hline\hline
		&\multicolumn{2}{c}{X-ray Source\sS{b}}	&\multicolumn{5}{c}{Stellar Model A\sS{c}} 						&{Stellar Model B\sS{d}}	\\	
\cline{2-3}
\cline{4-8}
		&Surface 	&Volume			&Star volume 	&Star surface	&X-ray to 	 &X-ray		&Required CV	& Star volume			\\
Field		&density 	&density		&density	&mass density	& stellar mass	&to stars	&to stars\sS{e}	& density			\\
		& (deg\sS{-2})	&(10\sS{-7} pc\sS{-3})	& (pc\sS{-3}) 	&(10\sS{8}\Ms deg\sS{-2})&(10\sS{-7}\Msi) &(10\sS{-7})	&(10\sS{-5})	& (pc\sS{-3}) 		\\
\hline
\multicolumn{3}{l}{Field to field Comparison}	\\
BW       	& 25(45)        & 3.2(5.7)      	& 2.1   	& 0.5		&5.1(9.1)      & 1.6(2.8)      & 2.6(4.6)           & 3.2 \\
SW       	& 39(47)        & 4.9(6.0)      	& 5.3   	& 1.3		&3.1(3.7)      & 0.9(1.1)      & 1.6(1.9)           & 6.7 \\
LW       	& 203(64)       & 25(8.1)       	& 7.6   	& 1.8		&11(3.5)       & 3.4(1.1)      & 5.6(1.8)           & 8.3 \\
Sgr B2   	& 582(93)       & 73(11)        	& 41    	& 10		&5.7(0.9)      & 1.8(0.3)      & 2.9(0.5)           & 8.8 \\
Sgr C    	& 902(113)	& 113(14)       	& 45    	& 11		&8.1(1.0)      & 2.5(0.3)      & 4.1(0.5)           & \sout{\it 8.9} \\
Arches   	& 1624(153)     & 204(19)       	& 52    	& 13		&12(1.2)       & 3.9(0.4)      & 6.5(0.6)           & \sout{\it 9.1} \\
Stacked Sgr A*  & 3100(187)     & 389(23)       	& 68    	& 17		&18(1.1)       & 5.7(0.3)      & 9.5(0.6)           & \sout{\it 9.1} \\
100 ksSgr A*    & 2117(176)     & 266(22)       	& 68   	 	& 17		&12(1.1)       & 3.9(0.3)      & 6.5(0.5)           & \sout{\it 9.1} \\
\hline
\multicolumn{4}{l}{By the fit to the radial distribution\sS{f}}	\\
\multicolumn{2}{l}{with stacked Sgr A* (Fig.~5b or 6a)}  	&      	& 		&             &  & 10.3(0.5)     & 3.2(0.2)      & 5.3(0.3)      	\\
\multicolumn{2}{l}{with 100ks Sgr A* (Fig.~6b)}    		&      	&		&             &  & 9.2(0.5)      & 2.8(0.2)      & 4.7(0.3)      	\\
\hline
\multicolumn{4}{l}{By the fit with the freed relative normalization of $\rho_A$\sS{g}}	\\
\multicolumn{2}{l}{with stacked Sgr A* (Fig.~6c)}      	&  &             &		&  & 7.1(0.6)      & 2.2(0.2)      & 3.6(0.3)      	\\
\multicolumn{2}{l}{with 100ks Sgr A* (Fig.~6d)}        	&  &             &		&  & 7.5(0.6)      & 2.3(0.2)      & 3.8(0.3)      	\\
\hline
\end{tabular*}
\\(a) Assumes the hard X-ray sources ($S> 1.5 \times 10^{-14}$ \fcgs in
\Hc) or stars seen within the
	$17\arcmin \times 17\arcmin$ FoV are 
	mainly ($\gtrsim$ 80\%) from 6--10 kpc distance.
(b) Using the PL+Fe model. We subtract the
expected AGN numbers, 145 deg\sS{-2} from the Window fields and 107 deg\sS{-2} from the
high extinction fields.
(c) The composite model in M06 and references therein.
The model gives the stellar mass density in the unit of \Ms pc\sS{-3}, and we assume the
local value of 0.144 stars pc\sS{-3} and 0.04 \Ms pc\sS{-3} to get
the star number density \citep{Robin03}. This relation should be good for the bulge in the case
of CVs and active binaries, but perhaps not good for young stars \citep{Sazonov06}.
(d) The stellar density model by \citet{Picaud04} for the Galactic disk
and outer Galactic bulge. The model does not include a central nucleus, so 
the (italic) values for the Sgr B2, Sgr C, Arches, and Sgr A* fields
are not reliable. See \S\ref{s:density}.
(e) The required CV to star density to explain the hard GB X-ray
sources by IPs. We assume that IPs are 5\% of
all CVs (e.g. $\sim$ 2--8\% in R06)
and that about 12\% of them
are detected above $10^{32}$ \lcgs (e.g.
$\sim$ 10--16\% in R06), which corresponds $\sim 1.5 \times
10^{-14}$ \fcgs for the sources near the GC (see text).
(f) The relative normalizations among stellar model components are
fixed as given in Section 4.3.
(g) The relative normalization of the central spherical cluster
component of the stellar distribution is allowed as a free parameter
and fitted as well.
\label{t:d}
\end{table*}

\begin{table*}
\small
\caption{Total X-ray and IR surface brightness (see also Fig.~\ref{f:surface})}
\begin{tabular*}{0.98\textwidth}{l@{\extracolsep{\fill}}cccccccccc}
\hline\hline
		&\multicolumn{2}{c}{X-ray\sS{a}	}			& \multicolumn{3}{c}{IR	}	\\
Field		&\multicolumn{2}{c}{2--8 keV (the \Hc band)}		& \multicolumn{3}{c}{3.6 $\mu$m}	\\
		&\multicolumn{2}{c}{(10\sS{-10} \fcgs deg\sS{-2}) }	& \multicolumn{3}{c}{(MJy/sr)}\\
\cline{2-3}
\cline{4-6}
		& PL		& PL+Fe					& observed	& unabsorbed (PL)\sS{b}	& unabsorbed (PL+Fe)\sS{c} \\
\hline

BW       	&0.34(0.02)	&0.35(0.02)				& 7.4(0.4)	& 8.2(0.4)  	& 8.2(0.4)		\\
SW       	&0.48(0.01)	&0.49(0.01)				& 15(1)		& 17(2)		& 17(2)			\\
LW       	&1.10(0.01)	&1.13(0.01)				& 18(1)		& 24(1)		& 24(1)			\\
Sgr B2   	&5.09(0.05)	&5.42(0.05)				& 26(6)		& 74(18)	& 150(36) 		\\
Sgr C    	&14.9(0.1)	&16.6(0.1)					& 35(3)		& 152(13)	& 384(33)      		\\
Arches   	&30.9(0.1)	&33.4(0.1)					& 58(12)	& 199(40)	& 337(69)		\\
Sgr A*  	&23.93(0.02)	&27.72(0.02)					& 68(16)	& 493(110)	& 1100(250)		\\
\hline
\end{tabular*}
(a) Assume the spectral model using the \Gm group for the Windows field and the \Gh group for the rest in Table~\ref{t:q}.
The error is of the statistical origin. Note the expected cosmic X-ray
background (CXB) is $(1.6-1.7 \pm 0.2) \times 10^{-11}$  \fcgs deg\sS{-2} for the Windows fields 
and $(1.3-1.4 \pm 0.2) \times 10^{-11}$ \fcgs deg\sS{-2} for the rest \citep{Hickox06}. See the \S\ref{s:surface}.
(b) The \nH estimate by PL (line B in Fig.~\ref{f:surface}). (c) The \nH estimate by PL+Fe (line A in Fig.~\ref{f:surface}).
\label{t:surface}
\end{table*}

\begin{table*}
\scriptsize
\caption{Specific Luminosity of X-ray point sources}
\begin{tabular*}{\textwidth}{l@{\extracolsep{\fill}}ccccll}
\hline\hline
			&				& Energy	&Luminosity		& Scaled for			&					 \\
Source			&Reported			& Range		&Range			& $10^{32.7-34}$ \lcgs in 2--8 keV	&Studied Fields					 \\
			& ($10^{26}$ \lcgs \Msi)	& (keV)		& (\lcgs)		& ($10^{26}$ \lcgs \Msi)		&							\\
\hline
This Study		& 0.5 -- 2.8			& 2--8 		&$10^{32.7-34}$ 	&  0.5 -- 2.8				&7 GB fields (100 or 750 ks)				\\
M06 			& 5.0 $\pm$ 2			& 0.5--8 	&$10^{32.7-34}$ 	&  3.3 $\pm$ 1.3 			&$2\Deg \times 1\Deg$ around the GC (100 or $2\times 12$ ks )			\\
S06			& 45 $\pm$ 9			& 2--10   	&$10^{27-36}$ 		&  9 $\pm$ 3 \sS{a} 			&the local solar neighborhood				\\
			& 				&    		& 			&  6 $\pm$ 2 \sS{b}			&							\\
R07b			& 77 $\pm$ 39			& 2--10 	&$10^{30.3-32.3}$ 	&  3 $\pm$ 1 \sS{c}			&the Sgr A* field (1 Ms)				\\
\hline
\end{tabular*}
M06 -- \citet{Muno06}, S06 -- \citet{Sazonov06}, R07b - \citet{Revnivtsev07b}\\
(a) by the scaling by M06. (b) Fig.~9 in S06. (c) assuming 3\% of the total emission 
($(4\pm 2) \times 10^{27}$ \lcgs \Msi) is from $> 10^{32.7}$ \lcgs, based on Fig.~6 in R07b.
\label{t:sl}
\end{table*}
\clearpage

\end{document}